\documentclass[notitlepage,a4paper,aps,prd,onecolumn,superscriptaddress,nofootinbib,longbibliography]{revtex4-1}
\usepackage{amsmath}
\usepackage{amsfonts}
\usepackage{amssymb}
\usepackage[latin1,utf8]{inputenc}
\usepackage[T1]{fontenc}
\usepackage[colorlinks=true]{hyperref}

\newcommand{\tp}[1]{\overset{\bullet}{#1}\vphantom{#1}}
\newcommand{\lc}[1]{\overset{\circ}{#1}\vphantom{#1}}

\newcommand{\dd}{\mathrm{d}}

\begin{document}

\title{Scalar-torsion theories of gravity III:\\analogue of scalar-tensor gravity and conformal invariants}

\author{Manuel Hohmann}
\email{manuel.hohmann@ut.ee}
\affiliation{Laboratory of Theoretical Physics, Institute of Physics, University of Tartu, W. Ostwaldi 1, 50411 Tartu, Estonia}

\begin{abstract}
We discuss a class of teleparallel scalar-torsion theories of gravity, which is parametrized by five free functions of the scalar field. The theories are formulated covariantly using a flat, but non-vanishing spin connection. We show how the actions of different theories within this class are related via conformal transformations of the tetrad and redefinitions of the scalar field, and derive the corresponding transformation laws for the free function in the action. From these we construct a number of quantities which are invariant under these transformations, and use them to write the action and field equations in different conformal frames. These results generalize a similar formalism for scalar-tensor theories of gravity, where the invariants have been used to express observables independently of the conformal frame.
\end{abstract}

\maketitle

\section{Introduction}\label{sec:intro}
An important and well-studied class of gravity theories, which have been used to address cosmological observations such as the accelerating expansion of the Universe at present and early times in its history, is given by scalar-tensor gravity theories~\cite{Faraoni:2004pi,Fujii:2003pa}. These theories have in common that they contain one or more scalar fields, which in general is non-minimally coupled to the metric of spacetime. The gravitational dynamics of the theory is then determined through the curvature of the Levi-Civita connection of the metric, as well as the dynamics of the scalar fields. A class of such theories of particular interest is defined in terms of four free functions in the action functional, where any specific choice of these functions defines a concrete theory~\cite{Flanagan:2004bz}.

An curious property of the aforementioned class of scalar-tensor theories is their behavior under conformal transformations. It has been shown that said transformations constitute maps between different theories within this class~\cite{Flanagan:2004bz}. It is an ongoing debate whether these conformally related theories are equivalent in their physical predictions~\cite{Catena:2006bd,Faraoni:2006fx,Deruelle:2010ht,Chiba:2013mha,Postma:2014vaa,Faraoni:1998qx,Capozziello:2010sc,Rondeau:2017xck}. An important contribution to this debate is the definition of a number of invariant quantities, which can then be used to express physical observables such that they become independent of the choice of the conformal frame~\cite{Jarv:2014hma,Kuusk:2015dda}.

Another thoroughly studied class of gravity theories is given by teleparallel models of gravity, where the gravitational interaction is attributed not to the curvature of the Levi-Civita connection, but to the torsion of a flat connection~\cite{Moller:1961,Aldrovandi:2013wha,Maluf:2013gaa,Golovnev:2018red}. The underlying teleparallel geometry provides another possible starting point for constructing new gravity theories by coupling scalar fields to torsion, and a number of such models have been studied~\cite{Geng:2011aj,Izumi:2013dca,Chakrabarti:2017moe,Otalora:2013tba,Jamil:2012vb,Chen:2014qsa}, as well as the question of conformal transformations. However, in these studies it is conventional to assume a fixed, vanishing spin connection, which potentially leads to the issue of local Lorentz symmetry breaking~\cite{Li:2010cg,Sotiriou:2010mv}, as spurious degrees of freedom may appear~\cite{Li:2011rn,Ong:2013qja,Izumi:2013dca,Chen:2014qtl}, and only recently the covariant formulation of teleparallel gravity~\cite{Krssak:2015oua} has been adopted to scalar-torsion gravity~\cite{Hohmann:2018rwf}.

The aim of our work is to combine several aspects of the aforementioned studies. We study a class of teleparallel scalar-torsion theories of gravity in the covariant formulation, which is constructed in analogy to the aforementioned class of scalar-(curvature)-tensor gravity theories, and contains scalar-tensor gravity as a subclass. Any specific theory of this class is determined by a particular choice of five free functions of the scalar field. We study the behavior of these theories under conformal transformations of the underlying teleparallel geometry, and show that such transformations relate different theories to each other. We then show that such classes of conformally related theories can be characterized by a number of invariant quantities, in full analogy to their scalar-tensor counterparts, and use these to define particular conformal frames.

This article belongs to a series of three articles on teleparallel scalar-torsion theories of gravity in the covariant formulation. In the first article~\cite{Hohmann:2018vle} we discussed the most general class of theories in which a scalar field is coupled to the tetrad and spin connection of teleparallel gravity, with the only restriction that the action is invariant under local Lorentz transformations and the matter fields do not couple to the spin connection (while allowing a coupling to the scalar field). The results derived in the first article were then used and applied to a particular subclass of scalar-torsion theories, which we called \(L(T, X, Y, \phi)\) theories, in the second article~\cite{Hohmann:2018dqh}. The class of theories we discuss in this article is a further restriction of the aforementioned class of \(L(T, X, Y, \phi)\) theories.

The outline of this article is as follows. In section~\ref{sec:action} we briefly review the dynamical fields of scalar-torsion theory based on teleparallel geometry and define the class of theories we consider here by giving their action functionals. The field equations for this class of theories are shown in section~\ref{sec:feqs}. We then turn our focus to conformal transformations. In section~\ref{sec:conformal} we derive how conformal transformations and scalar field redefinitions act on the scalar-torsion action. We then identify a set of invariant quantities under these transformations in section~\ref{sec:invariant}. These are used to define particular conformal frames in section~\ref{sec:frames}. In section~\ref{sec:multi} we show how these results can be generalized to multiple scalar fields. Specific examples are shown in section~\ref{sec:examples}, in particular the relation to scalar-tensor theories of gravity. We end with a conclusion in section~\ref{sec:conclusion}.

\section{Dynamical fields and action}\label{sec:action}
We start our discussion by introducing the dynamical fields and for the class of teleparallel scalar-torsion theories we consider in this article. Similarly to our previous work~\cite{Hohmann:2018vle,Hohmann:2018dqh} the dynamical fields are given by a coframe field \(\theta^a = \theta^a{}_{\mu}\dd x^{\mu}\), a flat spin connection \(\tp{\omega}^a{}_b = \tp{\omega}^a{}_{b\mu}\dd x^{\mu}\) and a scalar field \(\phi\). The frame field dual to the coframe field \(\theta^a\) will be denoted \(e_a = e_a{}^{\mu}\partial_{\mu}\). We denote quantities related to the flat spin connection with a bullet (\(\bullet\)). This in particular applies to the torsion tensor
\begin{equation}
T^{\rho}{}_{\mu\nu} = e_a{}^{\rho}\left(\partial_{\mu}e^a{}_{\nu} - \partial_{\nu}e^a{}_{\mu} + \tp{\omega}^a{}_{b\mu}e^b{}_{\nu} - \tp{\omega}^a{}_{b\nu}e^b{}_{\mu}\right)\,,
\end{equation}
the superpotential
\begin{equation}
S_{\rho\mu\nu} = \frac{1}{2}\left(T_{\nu\mu\rho} + T_{\rho\mu\nu} - T_{\mu\nu\rho}\right) - g_{\rho\mu}T^{\sigma}{}_{\sigma\nu} + g_{\rho\nu}T^{\sigma}{}_{\sigma\mu}
\end{equation}
and the torsion scalar
\begin{equation}\label{eqn:torsscal}
T = \frac{1}{2}T^{\rho}{}_{\mu\nu}S_{\rho}{}^{\mu\nu}\,.
\end{equation}
Here we made use of the metric
\begin{equation}\label{eqn:metric}
g_{\mu\nu} = \eta_{ab}\theta^a{}_{\mu}\theta^b{}_{\nu}\,,
\end{equation}
where \(\eta_{ab} = \mathrm{diag}(-1,1,1,1)\) is the Minkowski metric. Quantities associated to the Levi-Civita connection \(\lc{\nabla}_{\mu}\) will be denoted with an open circle (\(\circ\)). Further, we define the scalar field kinetic term
\begin{equation}\label{eqn:defx}
X = -\frac{1}{2}g^{\mu\nu}\phi_{,\mu}\phi_{,\nu}\,,
\end{equation}
as well as the derivative coupling term
\begin{equation}\label{eqn:defy}
Y = g^{\mu\nu}T^{\rho}{}_{\rho\mu}\phi_{,\nu}\,,
\end{equation}
which will enter the gravitational action introduced below.

The class of scalar-torsion theories we consider in this article has been studied, e.g., in the context of \(f(T,B)\) theories~\cite{Wright:2016ayu}, where
\begin{equation}\label{eqn:boundary}
B = \lc{R} + T = 2\lc{\nabla}_{\nu}T_{\mu}{}^{\mu\nu}\,.
\end{equation}
The gravitational part of the action we use here is given by
\begin{equation}\label{eqn:classactiong}
S_g\left[\theta^a, \tp{\omega}^a{}_b, \phi\right] = \frac{1}{2\kappa^2}\int_M\left[-\mathcal{A}(\phi)T + 2\mathcal{B}(\phi)X + 2\mathcal{C}(\phi)Y - 2\kappa^2\mathcal{V}(\phi)\right]\theta\dd^4x\,,
\end{equation}
where \(\mathcal{A}, \mathcal{B}, \mathcal{C}, \mathcal{V}\) are free functions of the scalar field and \(\theta = \det\theta^a{}_{\mu}\) is the volume element of the tetrad. Note that the action is reminiscent of scalar-tensor gravity, where a similar class of actions may be considered~\cite{Flanagan:2004bz}. This similarity is not by accident, and we will explore it further in section~\ref{ssec:stg}. One immediately sees that this action is of the form
\begin{equation}\label{eqn:confaction}
S_g\left[\theta^a, \tp{\omega}^a{}_b, \phi\right] = \int_ML\left(T, X, Y, \phi\right)\theta\dd^4x\,,
\end{equation}
where the Lagrangian is given by
\begin{equation}\label{eqn:actionsrels}
L = \frac{1}{2\kappa^2}\left[-\mathcal{A}(\phi)T + 2\mathcal{B}(\phi)X + 2\mathcal{C}(\phi)Y\right] - \mathcal{V}(\phi)\,.
\end{equation}
This class of actions has been studied in our previous work~\cite{Hohmann:2018dqh}, and it follows that all results derived therein also apply to the theories we study in this article. We will make use of this relation in the following section for deriving the field equations.

We further remark that alternatively we could study the action
\begin{equation}\label{eqn:classactionb}
S_g\left[\theta^a, \tp{\omega}^a{}_b, \phi\right] = \frac{1}{2\kappa^2}\int_M\left[-\mathcal{A}(\phi)T + 2\mathcal{B}(\phi)X - \tilde{\mathcal{C}}(\phi)B - 2\kappa^2\mathcal{V}(\phi)\right]\theta\dd^4x\,,
\end{equation}
which is equivalent to the action~\eqref{eqn:classactiong} for \(\mathcal{C} = \tilde{\mathcal{C}}'\), up to a boundary term. However, we will not do so for two reasons. First, the action~\eqref{eqn:classactionb} allows for an arbitrary shift \(\tilde{\mathcal{C}} \mapsto \tilde{\mathcal{C}} + \tilde{\mathcal{C}}_0\) of the function \(\tilde{\mathcal{C}}\) by a constant \(\tilde{\mathcal{C}}_0\), which changes the action by a boundary term, and hence does not alter the field equations. This arbitrariness is not present in the action~\eqref{eqn:classactiong}. Further, we will see in section~\ref{sec:multi} that the action~\eqref{eqn:classactiong} allows for a larger class of generalizations to multiple scalar fields, which affects the possibilities to choose particular conformal frames.

In addition to the gravitational part of the action, we now define a matter part. Also in analogy to scalar-tensor gravity we consider a matter coupling to a conformally rescaled tetrad, such that the matter action is of the form
\begin{equation}\label{eqn:classactionm}
S_m[\theta^a, \phi, \chi^I] = S_m^{\mathfrak{J}}\left[e^{\alpha(\phi)}\theta^a, \chi^I\right] = S_m^{\mathfrak{J}}\left[\theta^{\mathfrak{J}\,a}, \chi^I\right]\,,
\end{equation}
with another free function \(\alpha\) of the scalar field, and where we defined \(\theta^{\mathfrak{J}\,a} = e^{\alpha(\phi)}\theta^a\). The notation involving a superscript \(\mathfrak{J}\) will be explained in section~\ref{ssec:jordan}. For the variation of the matter action we write
\begin{equation}\label{eqn:matactvar}
\delta S_m[\theta^a, \phi, \chi^I] = \int_M\left(\Theta_a{}^{\mu}\delta\theta^a{}_{\mu} + \vartheta\delta\phi + \varpi_I\delta\chi^I\right)\theta\dd^4x\,.
\end{equation}
It follows from the structure~\eqref{eqn:classactionm} of the matter action that its variation can also be written as
\begin{equation}
\delta S_m^{\mathfrak{J}}\left[\theta^{\mathfrak{J}\,a}, \chi^I\right] = \int_M\left(\Theta^{\mathfrak{J}}_a{}^{\mu}\delta\theta^{\mathfrak{J}\,a}{}_{\mu} + \varpi^{\mathfrak{J}}_I\delta\chi^I\right)\theta^{\mathfrak{J}}\dd^4x = \int_M\left[\Theta^{\mathfrak{J}}_a{}^{\mu}e^{\alpha}\left(\delta\theta^a{}_{\mu} + \alpha'\theta^a{}_{\mu}\delta\phi\right) + \varpi^{\mathfrak{J}}_I\delta\chi^I\right]\theta^{\mathfrak{J}}\dd^4x\,.
\end{equation}
By comparing with the general variation~\eqref{eqn:matactvar} of the matter action we find that the matter terms \(\Theta_a{}^{\mu}\) and \(\vartheta\), which appear as coefficients of the variations \(\delta\theta^a{}_{\mu}\) and \(\delta\phi\) and which will enter the scalar and tetrad field equations, are related by
\begin{equation}\label{eqn:classenmomtens}
\vartheta = \alpha'\theta^a{}_{\mu}\Theta_a{}^{\mu} = \alpha'\Theta\,.
\end{equation}
We will make use of this relation when we display the field equations. These will be discussed in the following section.

\section{Field equations}\label{sec:feqs}
We now come to the field equations for the class of scalar-torsion theories introduced in the previous section, which are derived from the action~\eqref{eqn:classactiong} and~\eqref{eqn:classactionm}. For brevity, we will not display the full derivation of the field equations here, but make use of the relation~\eqref{eqn:actionsrels} to the class of theories defined by the action~\eqref{eqn:confaction}, whose field equations have been derived explicitly in our previous work~\cite{Hohmann:2018dqh}.

It follows from the structure of the dynamical fields that there are field equations derived by variations of the tetrad, the flat spin connection and the scalar field. However, it follows from the local Lorentz invariance of the action that the connection field equations are identical to the antisymmetric part of the tetrad field equations, so that the spin connection becomes a pure gauge degree of freedom, and only the symmetric part of the tetrad field equations remains independent; see~\cite{Hohmann:2018vle} for a detailed discussion. Here we make use of this fact and display only the independent parts of the field equations. For this purpose we compare the action with that of the more general \(L(T, X, Y, \phi)\) theory~\cite{Hohmann:2018dqh} and derive the terms
\begin{equation}
L_T = -\frac{\mathcal{A}(\phi)}{2\kappa^2}\,, \quad L_X = \frac{\mathcal{B}(\phi)}{\kappa^2}\,, \quad L_Y = \frac{\mathcal{C}(\phi)}{\kappa^2}\,, \quad L_{\phi} = \frac{1}{2\kappa^2}\left[-\mathcal{A}'(\phi)T + 2\mathcal{B}'(\phi)X + 2\mathcal{C}'(\phi)Y\right] - \mathcal{V}'(\phi)\,.
\end{equation}
which enter the gravitational field equations. We start with the symmetric part of the tetrad field equations, which take the form
\begin{multline}
\mathcal{A}'S_{(\mu\nu)}{}^{\rho}\phi_{,\rho} + \frac{1}{2}\mathcal{A}\left(2\lc{\nabla}_{\rho}S_{(\mu\nu)}{}^{\rho} - S_{(\mu}{}^{\rho\sigma}T_{\nu)\rho\sigma} + Tg_{\mu\nu}\right) + \left(\frac{1}{2}\mathcal{B} - \mathcal{C}'\right)\phi_{,\rho}\phi_{,\sigma}g^{\rho\sigma}g_{\mu\nu}\\
- (\mathcal{B} - \mathcal{C}')\phi_{,\mu}\phi_{,\nu} + \mathcal{C}\left(S_{(\mu\nu)}{}^{\rho}\phi_{,\rho} + \lc{\nabla}_{\mu}\lc{\nabla}_{\nu}\phi - \lc{\square}\phi g_{\mu\nu}\right) + \kappa^2\mathcal{V}g_{\mu\nu} = \kappa^2\Theta_{\mu\nu}\,,
\end{multline}
where we used the fact that
\begin{equation}
S_{(\mu\nu)}{}^{\rho}\phi_{,\rho} = T_{(\mu\nu)}{}^{\rho}\phi_{,\rho} + T^{\rho}{}_{\rho(\mu}\phi_{,\nu)} - T_{\rho}{}^{\rho\sigma}\phi_{,\sigma}g_{\mu\nu}\,.
\end{equation}
We can further simplify this expression using the identity
\begin{equation}
\lc{\nabla}_{\rho}S_{(\mu\nu)}{}^{\rho} - \frac{1}{2}S_{(\mu}{}^{\rho\sigma}T_{\nu)\rho\sigma} + \frac{1}{2}Tg_{\mu\nu} = \lc{R}_{\mu\nu} - \frac{1}{2}\lc{R}g_{\mu\nu}
\end{equation}
for the Einstein tensor, such that the symmetric part of the tetrad field equations finally reads
\begin{multline}\label{eqn:clafeqtets}
\left(\mathcal{A}' + \mathcal{C}\right)S_{(\mu\nu)}{}^{\rho}\phi_{,\rho} + \mathcal{A}\left(\lc{R}_{\mu\nu} - \frac{1}{2}\lc{R}g_{\mu\nu}\right) + \left(\frac{1}{2}\mathcal{B} - \mathcal{C}'\right)\phi_{,\rho}\phi_{,\sigma}g^{\rho\sigma}g_{\mu\nu}\\
- (\mathcal{B} - \mathcal{C}')\phi_{,\mu}\phi_{,\nu} + \mathcal{C}\left(\lc{\nabla}_{\mu}\lc{\nabla}_{\nu}\phi - \lc{\square}\phi g_{\mu\nu}\right) + \kappa^2\mathcal{V}g_{\mu\nu} = \kappa^2\Theta_{\mu\nu}\,.
\end{multline}
The antisymmetric part of the tetrad field equations, which is identical to the connection field equations, is given by
\begin{equation}\label{eqn:clafeqcon}
(\mathcal{A}' + \mathcal{C})T^{\rho}{}_{[\mu\nu}\phi_{,\rho]} = 0\,.
\end{equation}
Finally, the scalar field equation takes the form
\begin{equation}\label{eqn:clafeqscal}
\frac{1}{2}\mathcal{A}'T - \mathcal{B}\lc{\square}\phi - \frac{1}{2}\mathcal{B}'g^{\mu\nu}\phi_{,\mu}\phi_{,\nu} + \mathcal{C}\lc{\nabla}_{\mu}T_{\nu}{}^{\nu\mu} + \kappa^2\mathcal{V}' = \kappa^2\alpha'\Theta\,,
\end{equation}
where the trace \(\Theta = \theta^a{}_{\mu}\Theta_a{}^{\mu} = g^{\mu\nu}\Theta_{\mu\nu}\) of the energy-momentum tensor enters the scalar field equation through the relation~\eqref{eqn:classenmomtens}. These are the field equations for the class of theories defined by the action~\eqref{eqn:classactiong} and~\eqref{eqn:classactionm}.

If one naively tries to solve these field equations one encounters the difficulty that the scalar field equation~\eqref{eqn:clafeqscal} contains second derivatives of both the tetrad and the scalar field. In order to find solutions, it is more convenient to remove the second derivatives of the tetrad by a suitable linear combination of the tetrad field equations; this procedure is also called ``debraiding''~\cite{Bettoni:2015wta}. Using the identity \(S_{\mu}{}^{\mu\nu} = -2T_{\mu}{}^{\mu\nu}\), we take the trace
\begin{equation}
-2\left(\mathcal{A}' + \mathcal{C}\right)T_{\mu}{}^{\mu\nu}\phi_{,\nu} - \mathcal{A}\lc{R} + \left(\mathcal{B} - 3\mathcal{C}'\right)g^{\mu\nu}\phi_{,\mu}\phi_{,\nu} - 3\mathcal{C}\lc{\square}\phi + 4\kappa^2\mathcal{V} = \kappa^2\Theta
\end{equation}
of the symmetric part~\eqref{eqn:clafeqtets}. Together with the relation~\eqref{eqn:boundary} we find the debraided scalar field equation
\begin{equation}\label{eqn:clafeqdeb}
(\mathcal{A}' + \mathcal{C})\left(\mathcal{A}T - 2\mathcal{C}T_{\mu}{}^{\mu\nu}\phi_{,\nu}\right) - \left(2\mathcal{A}\mathcal{B} + 3\mathcal{C}^2\right)\lc{\square}\phi + (\mathcal{B}\mathcal{C} - \mathcal{A}\mathcal{B}' - 3\mathcal{C}\mathcal{C}')g^{\mu\nu}\phi_{,\mu}\phi_{,\nu} + 2\kappa^2(\mathcal{A}\mathcal{V}' + 2\mathcal{C}\mathcal{V}) = \kappa^2(2\mathcal{A}\alpha' + \mathcal{C})\Theta\,.
\end{equation}
We see that the trace \(\Theta\) may act as the source of the scalar field through the coupling term \(Y\) in the gravitational action~\eqref{eqn:classactiong} also when the matter action~\eqref{eqn:classactionm} is independent of the scalar field. Hence, it is reasonable to say that the scalar field is minimally coupled when the debraided equation~\eqref{eqn:clafeqdeb} is source-free, \(2\mathcal{A}\alpha' + \mathcal{C} = 0\); otherwise, we call it non-minimally coupled.

The debraided scalar field equation~\eqref{eqn:clafeqdeb} contains no second derivatives of the tetrad. However, it is not possible to remove the second derivatives of the scalar field from the tetrad field equations~\eqref{eqn:clafeqtets} by the same procedure. In order to achieve a full debraiding of this type, one has to perform a conformal transformation to a particular frame. We will discuss conformal transformations in the following section, and show how this debraiding is done in section~\ref{ssec:debraiding}.

\section{Conformal transformations}\label{sec:conformal}
We now discuss the behavior of the action~\eqref{eqn:classactiong} and~\eqref{eqn:classactionm} introduced in section~\ref{sec:action} under conformal transformations of the tetrad and redefinitions of the scalar field. Under this type of transformation the dynamical variables change according to
\begin{equation}\label{eqn:conftrans}
\bar{\theta}^a{}_{\mu} = e^{\gamma(\phi)}\theta^a{}_{\mu}\,, \quad \bar{e}_a{}^{\mu} = e^{-\gamma(\phi)}e_a{}^{\mu}\,, \quad \bar{\phi} = f(\phi)\,,
\end{equation}
while the spin connection \(\tp{\omega}^a{}_b\) and matter variables \(\chi\) are not affected. As a consequence, also the terms in the gravitational part~\eqref{eqn:classactiong} of the action change according to the rules
\begin{equation}\label{eqn:scaltrans}
\bar{T} = e^{-2\gamma}\left(T + 4\gamma'Y + 12(\gamma')^2X\right)\,, \quad
\bar{Y} = e^{-2\gamma}f'(Y + 6\gamma'X)\,, \quad
\bar{X} = e^{-2\gamma}(f')^2X\,;
\end{equation}
see~\cite{Hohmann:2018dqh} for a more detailed derivation.

We then consider a different action functional \(\bar{S}\) with gravitational part \(\bar{S}_g\) and matter part \(\bar{S}_m\), which is obtained from the original action~\eqref{eqn:classactiong} and~\eqref{eqn:classactionm} by replacing the parameter functions \(\mathcal{A}, \mathcal{B}, \mathcal{C}, \mathcal{V}, \alpha\) with a new set of parameter functions \(\bar{\mathcal{A}}, \bar{\mathcal{B}}, \bar{\mathcal{C}}, \bar{\mathcal{V}}, \bar{\alpha}\). Evaluating this new action functional for the transformed fields \(\bar{\theta}^a\) and \(\bar{\phi}\) we find, making use of the relations~\eqref{eqn:conftrans} and in turn also~\eqref{eqn:scaltrans}, that the gravitational part \(\bar{S}_g\) of the new action satisfies
\begin{equation}\label{eqn:classacttransg}
\begin{split}
\bar{S}_g\left[\bar{\theta}^a, \tp{\omega}^a{}_b, \bar{\phi}\right] &= \frac{1}{2\kappa^2}\int_M\left[-\bar{\mathcal{A}}(\bar{\phi})\bar{T} + 2\bar{\mathcal{B}}(\bar{\phi})\bar{X} + 2\bar{\mathcal{C}}(\bar{\phi})\bar{Y} - 2\kappa^2\bar{\mathcal{V}}(\bar{\phi})\right]\bar{\theta}\dd^4x\\
&= \frac{1}{2\kappa^2}\int_M\bigg\{-e^{2\gamma(\phi)}\bar{\mathcal{A}}(f(\phi))T + 2e^{2\gamma(\phi)}\left[\bar{\mathcal{C}}(f(\phi))f'(\phi) - 2\bar{\mathcal{A}}\gamma'(\phi)\right]Y - 2\kappa^2e^{4\gamma(\phi)}\bar{\mathcal{V}}(f(\phi))\\
&\phantom{=}+ 2e^{2\gamma(\phi)}\left[\bar{\mathcal{B}}(f(\phi))f'^2(\phi) - 6\bar{\mathcal{A}}(f(\phi))\gamma'^2(\phi) + 6\bar{\mathcal{C}}(f(\phi))f'(\phi)\gamma'(\phi)\right]X\bigg\}\theta\dd^4x\,,
\end{split}
\end{equation}
while for its matter part \(\bar{S}_m\) holds
\begin{equation}\label{eqn:classacttransm}
\bar{S}_m\left[\bar{\theta}^a, \bar{\phi}, \chi^I\right] = S_m^{\mathfrak{J}}\left[e^{\bar{\alpha}(\bar{\phi})}\bar{\theta}^a, \chi^I\right] = S_m^{\mathfrak{J}}\left[e^{\bar{\alpha}(f(\phi)) + \gamma(\phi)}\theta^a, \chi^I\right]\,.
\end{equation}
By comparison to the original action~\eqref{eqn:classactiong} and~\eqref{eqn:classactionm} we find that the new action \(\bar{S}\), evaluated at the transformed (barred) fields, reproduces the original action \(S\), evaluated at the untransformed (unbarred) fields,
\begin{equation}
\bar{S}_g\left[\bar{\theta}^a, \tp{\omega}^a{}_b, \bar{\phi}\right] = S_g\left[\theta^a, \tp{\omega}^a{}_b, \phi\right]\,, \quad
\bar{S}_m\left[\bar{\theta}^a, \bar{\phi}, \chi^I\right] = S_m\left[\theta^a, \phi, \chi^I\right]\,,
\end{equation}
provided that the parameter functions of the two actions are related to each other by the rules
\begin{subequations}\label{eqn:pftrans}
\begin{align}
\mathcal{A} &= e^{2\gamma}\bar{\mathcal{A}}\,,\label{eqn:pftransA}\\
\mathcal{B} &= e^{2\gamma}\left(\bar{\mathcal{B}}f'^2 - 6\bar{\mathcal{A}}\gamma'^2 + 6\bar{\mathcal{C}}f'\gamma'\right)\,,\label{eqn:pftransB}\\
\mathcal{C} &= e^{2\gamma}\left(\bar{\mathcal{C}}f' - 2\bar{\mathcal{A}}\gamma'\right)\,,\label{eqn:pftransC}\\
\mathcal{V} &= e^{4\gamma}\bar{\mathcal{V}}\,,\label{eqn:pftransV}\\
\alpha &= \bar{\alpha} + \gamma\,.\label{eqn:pftransalpha}
\end{align}
\end{subequations}
Here we have omitted the function arguments for brevity; it is understood that transformed (barred) functions depend on \(\bar{\phi} = f(\phi)\), while all other (unbarred) functions depend on \(\phi\). Hence, we may say that the action functionals \(S\) and \(\bar{S}\), with parameter functions related by~\eqref{eqn:pftrans}, are related by the conformal transformation~\eqref{eqn:conftrans}. Since they are of the same form, we may also say that the transformation~\eqref{eqn:conftrans} preserves the form of the action.

We finally remark that the transformation of the matter action also induces a transformation of the matter terms in the field equations, which can be written in the form
\begin{equation}\label{eqn:mattermtrans}
\Theta_{\mu\nu} = e^{2\gamma}\bar{\Theta}_{\mu\nu}\,, \quad
\Theta = e^{4\gamma}\bar{\Theta}\,, \quad
\vartheta = e^{4\gamma}(\gamma'\bar{\Theta} + f'\bar{\vartheta})\,;
\end{equation}
see~\cite{Hohmann:2018dqh} for a detailed derivation. These relations will be used later, when we apply the conformal transformations to the field equations. Note also that the transformations~\eqref{eqn:mattermtrans}, together with the relation~\eqref{eqn:pftransalpha}, preserve the relation~\eqref{eqn:classenmomtens} in the sense that \(\bar{\theta} = \bar{\alpha}'\bar{\Theta}\).

One can see from the transformation behavior~\eqref{eqn:pftrans} of the parameter functions that there exist particular quantities constructed from these functions which transform trivially under conformal transformations. We will explicitly construct such quantities in the following section.

\section{Invariant quantities}\label{sec:invariant}
We have seen in the previous section that the class of theories we consider in this article exhibits a form invariance of their actions under conformal transformations of the tetrad and redefinitions of the scalar field. This form invariance and the corresponding transformation~\eqref{eqn:pftrans} of its constituting parameter functions \(\mathcal{A}, \mathcal{B}, \mathcal{C}, \mathcal{V}, \alpha\) is reminiscent of scalar-tensor gravity, where a similar transformation behavior can be found~\cite{Flanagan:2004bz}. In the latter class of theories it has motivated the introduction of a set of invariant functions~\cite{Jarv:2014hma}; these functions have subsequently been used to express a number of physical observables in a frame independent form~\cite{Jarv:2015kga,Kuusk:2016rso,Jarv:2016sow,Karam:2017zno}. We now show that the same type of invariants can also be introduced for the class of scalar-torsion theories we consider here, and we expect them to be of similar use for expressing physical observables independently of the choice of the conformal frame, as we will argue in more detail towards the end of this section.

From the transformation rules~\eqref{eqn:pftransA}, \eqref{eqn:pftransV} and~\eqref{eqn:pftransalpha} one can see immediately that the functions
\begin{equation}\label{eqn:cinv12}
\mathcal{I}_1 = \frac{e^{2\alpha}}{\mathcal{A}}\,, \quad
\mathcal{I}_2 = \frac{\mathcal{V}}{\mathcal{A}^2}
\end{equation}
are invariant under conformal transformations and scalar field redefinitions. Here invariance means that under a transformation of the form~\eqref{eqn:conftrans} they change according to
\begin{equation}\label{eqn:invtrans}
\bar{\mathcal{I}}_i(\bar{\phi}(x)) = \bar{\mathcal{I}}_i(f(\phi(x))) = \mathcal{I}_i(\phi(x))\,,
\end{equation}
which means that the functional forms of \(\mathcal{I}_i\) and \(\bar{\mathcal{I}}_i\) differ, but their values evaluated at each spacetime point \(x\) agree, provided that the scalar field is appropriately transformed, for \(i = 1, 2\). In contrast, the functions \(\mathcal{F}\) and \(\mathcal{H}\) defined by
\begin{equation}\label{eqn:cinvfh}
\mathcal{F} = \frac{2\mathcal{A}\mathcal{B} - 3\mathcal{A}'[2\mathcal{C} + \mathcal{A}']}{4\mathcal{A}^2}\,, \quad
\mathcal{H} = \frac{\mathcal{C} + \mathcal{A}'}{2\mathcal{A}}\,.
\end{equation}
are invariant under conformal transformations of the tetrad, but transform covariantly under redefinitions of the scalar field. This means that they incur an additional factor, and transform as
\begin{equation}\label{eqn:fhtrans}
\bar{\mathcal{F}}(\bar{\phi}) = \frac{1}{f'^2(\phi)}\mathcal{F}(\phi)\,, \quad
\bar{\mathcal{H}}(\bar{\phi}) = \frac{1}{f'(\phi)}\mathcal{H}(\phi)\,,
\end{equation}
as can be seen from the transformation rules~\eqref{eqn:pftransB} and~\eqref{eqn:pftransC}. The same behavior can be found also for the quantities
\begin{equation}\label{eqn:cinvgk}
\mathcal{G} = \frac{\mathcal{B} - 6\alpha'[\mathcal{C} + \alpha'\mathcal{A}]}{2e^{2\alpha}}\,, \quad
\mathcal{K} = \frac{\mathcal{C} + 2\alpha'\mathcal{A}}{2e^{2\alpha}}\,,
\end{equation}
i.e., they likewise transform as
\begin{equation}\label{eqn:gktrans}
\bar{\mathcal{G}}(\bar{\phi}) = \frac{1}{f'^2(\phi)}\mathcal{G}(\phi)\,, \quad
\bar{\mathcal{K}}(\bar{\phi}) = \frac{1}{f'(\phi)}\mathcal{K}(\phi)\,.
\end{equation}
They are related to the previously defined invariants by the relations
\begin{equation}\label{eqn:fhgk}
\mathcal{F} = \mathcal{I}_1\mathcal{G} + 3\frac{\mathcal{I}_1'}{\mathcal{I}_1}\left(\mathcal{I}_1\mathcal{K} - \frac{\mathcal{I}_1'}{4\mathcal{I}_1}\right)\,, \quad \mathcal{H} = \mathcal{I}_1\mathcal{K} - \frac{\mathcal{I}_1'}{2\mathcal{I}_1}\,.
\end{equation}
The invariant \(\mathcal{K}\) is closely related to the notion of minimal coupling we introduced at the end of section~\ref{sec:feqs}. We see that the scalar field is minimally coupled, i.e., the debraided field equation~\eqref{eqn:clafeqdeb} is source-free, if and only if \(\mathcal{K} = 0\). This condition is invariant under conformal transformations and scalar field redefinitions.

The are numerous possibilities to construct further invariants from those introduced above. For example, one may find quantities which are also invariant under scalar field redefinitions by taking the indefinite integrals
\begin{equation}
\int\sqrt{\mathcal{F}(\phi)}\dd\phi\,, \quad \int\mathcal{H}(\phi)\dd\phi\,,
\end{equation}
and similarly for \(\mathcal{G}\) and \(\mathcal{K}\). Also note that quotients \(\mathcal{I}_i'/\mathcal{I}_j'\) of invariants are again invariants, and that one may form invariant derivative operators; however, we will not pursue this direction further within the scope of this article, since these constructions are identical to those that may be defined in the case of scalar-tensor theories of gravity~\cite{Jarv:2014hma}. Instead, we will make use of the invariants to construct particular conformal frames, and derive expressions for the action functional and field equations which are invariant under conformal transformations. This will be done in the next section.

\section{Conformal frames}\label{sec:frames}
We have seen in section~\ref{sec:conformal} that under a conformal transformation of the tetrad and a redefinition of the scalar field of the form~\eqref{eqn:conftrans} the action~\eqref{eqn:classactiong} and~\eqref{eqn:classactionm} retains its form, provided that the defining functions of the scalar field are also transformed using the rules~\eqref{eqn:pftrans}. This freedom of transforming the action is also present in scalar-tensor theories of gravity, where it is commonly used to transform the action into two particular classes of parametrizations, known as Jordan and Einstein frames, in which the action and field equations exhibit additional properties. It has further been shown that these frames in scalar-tensor theories of gravity can be expressed in terms of a particular set of invariant quantities. We will now show that the same is possible also for the class of scalar-torsion theories we discuss in this article, making use of the invariants we defined in the preceding section.

We start by making use of the similarity to scalar-tensor gravity to define the Jordan frame in section~\ref{ssec:jordan} and the Einstein frame in section~\ref{ssec:einstein}. We will see that in contrast to scalar-tensor gravity, the naively defined Einstein frame does not to a complete debraiding of the scalar and tetrad field equations, as discussed at the end of section~\ref{sec:feqs}. However, we will define another frame in section~\ref{ssec:debraiding} in which this debraiding is obtained. Note that we will leave the scalar field unchanged in this section, \(\bar{\phi} = \phi\), unless otherwise noted.

\subsection{Jordan frame}\label{ssec:jordan}
We start with the Jordan frame, whose associated tetrad we define as
\begin{equation}\label{eqn:jframedef}
\theta^{\mathfrak{J}\,a} = e^{\gamma^{\mathfrak{J}}(\phi)}\theta^a = e^{\alpha(\phi)}\theta^a\,, \quad
\gamma^{\mathfrak{J}}(\phi) = \alpha(\phi)\,.
\end{equation}
It follows directly from this definition that the Jordan frame tetrad is invariant under conformal transformations and scalar field redefinitions of the original field variables in the sense that
\begin{equation}
\theta^{\mathfrak{J}\,a} = e^{\alpha(\phi)}\theta^a = e^{\bar{\alpha}(\bar{\phi}) + \gamma(\phi)}\theta^a = e^{\bar{\alpha}(\bar{\phi})}\bar{\theta}^a = \bar{\theta}^{\mathfrak{J}\,a}\,.
\end{equation}
Using the definition~\eqref{eqn:jframedef} for the function \(\gamma^{\mathfrak{J}}\), substituting it into the transformation rules~\eqref{eqn:pftrans} and comparing the obtained transformed (barred) parameter functions with the invariants detailed in section~\ref{sec:invariant}, we find the relations
\begin{equation}\label{eqn:pfjordan}
\mathcal{A}^{\mathfrak{J}} = \frac{1}{\mathcal{I}_1}\,, \quad
\mathcal{B}^{\mathfrak{J}} = 2\mathcal{G}\,, \quad
\mathcal{C}^{\mathfrak{J}} = 2\mathcal{K}\,, \quad
\mathcal{V}^{\mathfrak{J}} = \frac{\mathcal{I}_2}{\mathcal{I}_1^2}\,, \quad
\alpha^{\mathfrak{J}} = 0\,,
\end{equation}
where we have replaced the bars with superscripts \(\mathfrak{J}\), in order to indicate that this is the Jordan frame parametrization. The action can now be written in the form
\begin{equation}\label{eqn:jfaction}
S^{\mathfrak{J}}\left[\theta^{\mathfrak{J}\,a}, \tp{\omega}^a{}_b, \phi, \chi^I\right] = \frac{1}{2\kappa^2}\int_M\left[-\frac{1}{\mathcal{I}_1(\phi)}T^{\mathfrak{J}} + 4\mathcal{G}(\phi)X^{\mathfrak{J}} + 4\mathcal{K}(\phi)Y^{\mathfrak{J}} - 2\kappa^2\frac{\mathcal{I}_2(\phi)}{\mathcal{I}_1^2(\phi)}\right]\theta^{\mathfrak{J}}\dd^4x + S_m^{\mathfrak{J}}\left[\theta^{\mathfrak{J}\,a}, \chi^I\right]\,.
\end{equation}
A number of remarks are in order. First, note that the matter action functional in the Jordan frame action~\eqref{eqn:jfaction} agrees with the action functional we used in the definition~\eqref{eqn:classactionm} of the matter action; this is the reason for using the notation involving the superscript \(\mathfrak{J}\). Further, we see that \(S_m^{\mathfrak{J}}\) depends only on the Jordan frame tetrad and matter fields, and carries no additional, explicit dependence on the scalar field besides the implicit dependence through the definition~\eqref{eqn:jframedef}. This is the most important advantage and typical reason for using the Jordan frame, since also the resulting matter field equations \(\varpi^{\mathfrak{J}}_I = 0\) are expressed in terms of the Jordan frame tetrad and matter fields only, without further dependence on the scalar field. It further follows that the term \(\vartheta^{\mathfrak{J}}\) obtained from varying the matter action \(S_m^{\mathfrak{J}}\) with respect to the scalar field, while keeping the Jordan frame tetrad fixed, vanishes, and hence does not appear in the field equations, which we will show below.

We also remark that the gravitational part of the Jordan frame action~\eqref{eqn:jfaction} is defined only up to a redefinition of the scalar field. This means that we may we define a different Jordan frame action \(\bar{S}^{\mathfrak{J}}\) by replacing the invariant parameter functions \(\mathcal{I}_1, \mathcal{I}_2, \mathcal{G}, \mathcal{K}\) by their barred counterparts \(\bar{\mathcal{I}}_1, \bar{\mathcal{I}}_2, \bar{\mathcal{G}}, \bar{\mathcal{K}}\), which we then evaluate at the transformed scalar field,
\begin{equation}
\bar{S}^{\mathfrak{J}}\left[\theta^{\mathfrak{J}\,a}, \tp{\omega}^a{}_b, \bar{\phi}, \chi^I\right] = \frac{1}{2\kappa^2}\int_M\left[-\frac{1}{\bar{\mathcal{I}}_1(\bar{\phi})}T^{\mathfrak{J}} + 4\bar{\mathcal{G}}(\bar{\phi})\bar{X}^{\mathfrak{J}} + 4\bar{\mathcal{K}}(\bar{\phi})\bar{Y}^{\mathfrak{J}} - 2\kappa^2\frac{\bar{\mathcal{I}}_2(\bar{\phi})}{\bar{\mathcal{I}}_1^2(\bar{\phi})}\right]\theta^{\mathfrak{J}}\dd^4x + S_m^{\mathfrak{J}}\left[\theta^{\mathfrak{J}\,a}, \chi^I\right]\,.
\end{equation}
Substituting \(\bar{\phi} = f(\phi)\) we then find that the new action \(\bar{S}^{\mathfrak{J}}\), evaluated at \(\bar{\phi}\), agrees with the original action \(S^{\mathfrak{J}}\), evaluated at \(\phi\), provided that their defining parameter functions obey the transformation rules~\eqref{eqn:invtrans} and~\eqref{eqn:gktrans}. Note that we do not need to transform the matter part of the action~\eqref{eqn:jfaction} here, as it is independent of the scalar field.

We now express the field equations in the Jordan frame. The symmetric tetrad field equations
\begin{multline}\label{eqn:jfeqtets}
2\frac{\mathcal{H}}{\mathcal{I}_1}S^{\mathfrak{J}}_{(\mu\nu)}{}^{\rho}\phi_{,\rho} + \frac{1}{\mathcal{I}_1}\left(\lc{R}^{\mathfrak{J}}_{\mu\nu} - \frac{1}{2}\lc{R}^{\mathfrak{J}}g^{\mathfrak{J}}_{\mu\nu}\right) + \left(\mathcal{G} - 2\mathcal{K}'\right)\phi_{,\rho}\phi_{,\sigma}g^{\mathfrak{J}\,\rho\sigma}g^{\mathfrak{J}}_{\mu\nu}\\
- 2(\mathcal{G} - \mathcal{K}')\phi_{,\mu}\phi_{,\nu} + 2\mathcal{K}\left(\lc{\nabla}^{\mathfrak{J}}_{\mu}\lc{\nabla}^{\mathfrak{J}}_{\nu}\phi - \lc{\square}^{\mathfrak{J}}\phi g^{\mathfrak{J}}_{\mu\nu}\right) + \kappa^2\frac{\mathcal{I}_2}{\mathcal{I}_1^2}g^{\mathfrak{J}}_{\mu\nu} = \kappa^2\Theta^{\mathfrak{J}}_{\mu\nu}\,.
\end{multline}
and connection field equations
\begin{equation}\label{eqn:jfeqcon}
\mathcal{H}T^{\mathfrak{J}\,\rho}{}_{[\mu\nu}\phi_{,\rho]} = 0\,.
\end{equation}
are essentially unchanged compared to their general frame forms~\eqref{eqn:clafeqtets} and~\eqref{eqn:clafeqcon}, while the scalar field equation~\eqref{eqn:clafeqscal} becomes
\begin{equation}\label{eqn:jfeqscal}
-\frac{\mathcal{I}_1'}{2\mathcal{I}_1^2}T^{\mathfrak{J}} - 2\mathcal{G}\lc{\square}^{\mathfrak{J}}\phi - \mathcal{G}'g^{\mathfrak{J}\,\mu\nu}\phi_{,\mu}\phi_{,\nu} + 2\mathcal{K}\lc{\nabla}^{\mathfrak{J}}_{\mu}T^{\mathfrak{J}}_{\nu}{}^{\nu\mu} + \kappa^2\frac{\mathcal{I}_1\mathcal{I}_2' - 2\mathcal{I}_1\mathcal{I}_2'}{\mathcal{I}_1^3} = 0\,,
\end{equation}
and hence does not contain the matter energy-momentum tensor. Note, however, that the matter energy-momentum still acts as a source for the scalar field through the debraiding discussed at the end of section~\ref{sec:feqs}. This can be seen from the debraided scalar field equation~\eqref{eqn:clafeqdeb}, which reads
\begin{equation}\label{eqn:jfeqdeb}
2\frac{\mathcal{H}}{\mathcal{I}_1}\left(\frac{1}{\mathcal{I}_1}T^{\mathfrak{J}} + 2\mathcal{K}S^{\mathfrak{J}}_{\mu}{}^{\mu\nu}\phi_{,\nu}\right) - 4\frac{\mathcal{F} + 3\mathcal{H}^2}{\mathcal{I}_1^2}\lc{\square}^{\mathfrak{J}}\phi + \left[4\mathcal{K}(\mathcal{G} - 3\mathcal{K}') - 2\frac{\mathcal{G}'}{\mathcal{I}_1}\right]g^{\mathfrak{J}\,\mu\nu}\phi_{,\mu}\phi_{,\nu} + 2\kappa^2\frac{4\mathcal{H}\mathcal{I}_2 + \mathcal{I}_2'}{\mathcal{I}_1^3} = 2\kappa^2\mathcal{K}\Theta^{\mathfrak{J}}\,.
\end{equation}
in the Jordan frame.

\subsection{Einstein frame}\label{ssec:einstein}
We then come to the Einstein frame, which we construct following essentially the same procedure as for the Jordan frame above, but using the conformal transformation defined by
\begin{equation}\label{eqn:eframedef}
\theta^{\mathfrak{E}\,a} = e^{\gamma^{\mathfrak{E}}(\phi)}\theta^a = \sqrt{\mathcal{A}(\phi)}\theta^a\,, \quad
\gamma^{\mathfrak{E}}(\phi) = \frac{1}{2}\ln\mathcal{A}(\phi)\,.
\end{equation}
Similarly to the Jordan frame, also the Einstein frame tetrad is invariant under conformal transformations and scalar field redefinitions of the original field variables,
\begin{equation}
\theta^{\mathfrak{E}\,a} = \sqrt{\mathcal{A}(\phi)}\theta^a = \sqrt{\bar{\mathcal{A}}(\phi)}e^{\gamma(\phi)}\theta^a = \sqrt{\bar{\mathcal{A}}(\phi)}\bar{\theta}^a = \bar{\theta}^{\mathfrak{E}\,a}\,.
\end{equation}
Using the transformation rules~\eqref{eqn:pftrans} and the invariant quantities defined in section~\ref{sec:invariant}, we find that the parameter functions in the Einstein frame in terms of invariants are given by
\begin{equation}\label{eqn:pfeinstein}
\mathcal{A}^{\mathfrak{E}} = 1\,, \quad
\mathcal{B}^{\mathfrak{E}} = 2\mathcal{F}\,, \quad
\mathcal{C}^{\mathfrak{E}} = 2\mathcal{H}\,, \quad
\mathcal{V}^{\mathfrak{E}} = \mathcal{I}_2\,, \quad
\alpha^{\mathfrak{E}} = \frac{1}{2}\ln\mathcal{I}_1\,.
\end{equation}
In this case the action takes the form
\begin{equation}\label{eqn:efaction}
S^{\mathfrak{E}}\left[\theta^{\mathfrak{E}\,a}, \tp{\omega}^a{}_b, \phi, \chi^I\right] = \frac{1}{2\kappa^2}\int_M\left[-T^{\mathfrak{E}} + 4\mathcal{F}(\phi)X^{\mathfrak{E}} + 4\mathcal{H}(\phi)Y^{\mathfrak{E}} - 2\kappa^2\mathcal{I}_2(\phi)\right]\theta^{\mathfrak{E}}\dd^4x + S_m^{\mathfrak{J}}\left[\sqrt{\mathcal{I}_1(\phi)}\theta^{\mathfrak{E}\,a}, \chi^I\right]\,.
\end{equation}
Also in this case we add a few remarks. First, note that we have expressed the matter part of the action through the Jordan frame action functional \(S_m^{\mathfrak{J}}\). This is necessary in order to implement the particular relation between the dependences of the action on the tetrad and the scalar field imposed by the structure of the action~\eqref{eqn:classactionm}. We also see that in this case the matter action carries an explicit dependence on the scalar field, in addition to the implicit dependence incurred from the Einstein frame tetrad. In contrast, the scalar field does not appear in the term involving the torsion scalar \(T^{\mathfrak{E}}\). This is the characteristic property of the Einstein frame if one follows the analogy to scalar-tensor gravity, where the scalar field does not couple to the Ricci scalar \(\lc{R}^{\mathfrak{E}}\) in the Einstein frame.

We further remark that also in this case the action is uniquely defined only up to scalar field redefinitions, as is is also the case in the Jordan frame, i.e., if we define a new action \(\bar{S}^{\mathfrak{E}}\) such that
\begin{equation}
\bar{S}^{\mathfrak{E}}\left[\theta^{\mathfrak{E}\,a}, \tp{\omega}^a{}_b, \bar{\phi}, \chi^I\right] = \frac{1}{2\kappa^2}\int_M\left[-T^{\mathfrak{E}} + 4\bar{\mathcal{F}}(\bar{\phi})X^{\mathfrak{E}} + 4\bar{\mathcal{H}}(\bar{\phi})Y^{\mathfrak{E}} - 2\kappa^2\bar{\mathcal{I}}_2(\bar{\phi})\right]\theta^{\mathfrak{E}}\dd^4x + S_m^{\mathfrak{J}}\left[\sqrt{\bar{\mathcal{I}}_1(\bar{\phi})}\theta^{\mathfrak{E}\,a}, \chi^I\right]\,,
\end{equation}
and substitute the transformed scalar field \(\bar{\phi} = f(\phi)\), then we reproduce the original action~\eqref{eqn:efaction} for \(\phi\), provided that the invariant parameter functions satisfy the transformation rules~\eqref{eqn:invtrans} and~\eqref{eqn:fhtrans}. Also in this case the matter part \(S_m^{\mathfrak{J}}\) of the action is invariant, since \(\bar{\mathcal{I}}_1(\bar{\phi}) = \mathcal{I}_1(\phi)\) by the definition of the invariants.

Next, we come to the field equations. We find that the symmetric tetrad field equations~\eqref{eqn:clafeqtets} are given by
\begin{multline}\label{eqn:efeqtets}
2\mathcal{H}S^{\mathfrak{E}}_{(\mu\nu)}{}^{\rho}\phi_{,\rho} + \lc{R}^{\mathfrak{E}}_{\mu\nu} - \frac{1}{2}\lc{R}^{\mathfrak{E}}g^{\mathfrak{E}}_{\mu\nu} + \left(\mathcal{F} - 2\mathcal{H}'\right)\phi_{,\rho}\phi_{,\sigma}g^{\mathfrak{E}\,\rho\sigma}g_{\mu\nu}\\
- 2(\mathcal{F} - \mathcal{H}')\phi_{,\mu}\phi_{,\nu} + 2\mathcal{H}\left(\lc{\nabla}^{\mathfrak{E}}_{\mu}\lc{\nabla}^{\mathfrak{E}}_{\nu}\phi - \lc{\square}^{\mathfrak{E}}\phi g^{\mathfrak{E}}_{\mu\nu}\right) + \kappa^2\mathcal{I}_2g^{\mathfrak{E}}_{\mu\nu} = \kappa^2\Theta^{\mathfrak{E}}_{\mu\nu}\,.
\end{multline}
the connection field equations~\eqref{eqn:clafeqcon} read
\begin{equation}\label{eqn:efeqcon}
\mathcal{H}T^{\mathfrak{E}\,\rho}{}_{[\mu\nu}\phi_{,\rho]} = 0\,.
\end{equation}
and the scalar field equation~\eqref{eqn:clafeqscal} takes the form
\begin{equation}\label{eqn:efeqscal}
-2\mathcal{F}\lc{\square}^{\mathfrak{E}}\phi - \mathcal{F}'g^{\mathfrak{E}\,\mu\nu}\phi_{,\mu}\phi_{,\nu} + 2\mathcal{H}\lc{\nabla}^{\mathfrak{E}}_{\mu}T^{\mathfrak{E}}_{\nu}{}^{\nu\mu} + \kappa^2\mathcal{I}_2' = \kappa^2\alpha'\Theta^{\mathfrak{E}}\,,
\end{equation}
Finally, after debraiding we find the scalar field equation~\eqref{eqn:clafeqdeb} in the form
\begin{equation}\label{eqn:efeqdeb}
2\mathcal{H}\left(T^{\mathfrak{E}} + 2\mathcal{H}S^{\mathfrak{E}}_{\mu}{}^{\mu\nu}\phi_{,\nu}\right) - 4\left(\mathcal{F} + 3\mathcal{H}^2\right)\lc{\square}^{\mathfrak{E}}\phi + \left[4\mathcal{H}(\mathcal{F} - 3\mathcal{H}') - 2\mathcal{F}'\right]g^{\mathfrak{E}\,\mu\nu}\phi_{,\mu}\phi_{,\nu} + 2\kappa^2(4\mathcal{H}\mathcal{I}_2 + \mathcal{I}_2') = 2\kappa^2\mathcal{K}\mathcal{I}_1\Theta^{\mathfrak{E}}\,.
\end{equation}
From the symmetric part~\eqref{eqn:efeqtets} we see an important difference between scalar-tensor and scalar-torsion theories of gravity: in the scalar-tensor case there are no second derivatives of the scalar field in the metric field equation in the Einstein frame, leading to a complete debraiding of the metric and scalar field equations~\cite{Bettoni:2015wta}; this is not the case for the tetrad field equations of the class of scalar-torsion theories we discuss here, since the second order derivatives enter with a non-vanishing factor \(\mathcal{C}^{\mathfrak{E}} = 2\mathcal{H}\). Hence, the Einstein frame loses its debraiding property. One may argue that this fact renders the name Einstein frame questionable; we will comment on this below. Our choice to define the Einstein frame via \(\mathcal{A}^{\mathfrak{E}} = 1\) is motivated simply by its analogy to scalar-tensor gravity.

\subsection{Debraiding frame}\label{ssec:debraiding}
As we have seen above, the Einstein frame in the class of scalar-torsion gravity theories we consider in this article does not have the debraiding property that the field equations for the tetrad do not contain second derivatives of the scalar field. However, one can see from the structure of the field equations~\eqref{eqn:clafeqtets} that also in this case a debraiding can be achieved by performing a conformal transformation such that in the new frame, which we indicate by a superscript \(\mathfrak{D}\), the condition \(\mathcal{C}^{\mathfrak{D}} = 0\) is satisfied. By comparison with the transformation rule~\eqref{eqn:pftransC} we then find that this conformal transformation must satisfy
\begin{equation}
{\gamma^{\mathfrak{D}}}'(\phi) = -\frac{\mathcal{C}(\phi)}{2\mathcal{A}(\phi)}\,.
\end{equation}
Note that in contrast to the algebraic conditions~\eqref{eqn:jframedef} and~\eqref{eqn:eframedef} for the Jordan and Einstein frame transformations we obtain a differential equation, which fixes \(\gamma^{\mathfrak{D}}\) only up to an additive constant. Hence, also the corresponding debraiding tetrad \(\theta^{\mathfrak{D}\,a} = e^{\gamma^{\mathfrak{D}}(\phi)}\theta^a\) is determined only up to a constant factor. This could be fixed by the additional constraint that \(\gamma^{\mathfrak{D}}(\phi_0) = \gamma^{\mathfrak{D}}_0\) for some \(\phi_0\), such that
\begin{equation}
\gamma^{\mathfrak{D}}(\phi) = \gamma^{\mathfrak{D}}_0 - \frac{1}{2}\int_{\phi_0}^{\phi}\frac{\mathcal{C}(\tilde{\phi})}{\mathcal{A}(\tilde{\phi})}\dd\tilde{\phi}\,.
\end{equation}
However, this constraint would depend on the original frame, since also the frame transition function \(\gamma^{\mathfrak{D}}\) itself depends on the original frame. Hence, we will not follow this route. We will discuss other choices below, which do not have this problem.

Even without fixing the free constant factor in the definition of the debraiding tetrad \(\theta^{\mathfrak{D}\,a}\) one can determine the parameter functions in the debraiding frame up to a constant factor (or an additive constant in the case of \(\alpha^{\mathfrak{D}}\)). By comparison with the invariants introduced in section~\ref{sec:invariant} and using the condition \(\mathcal{C}^{\mathfrak{D}} = 0\) we find the relations
\begin{subequations}\label{eqn:pfdebraid}
\begin{align}
\left(\ln\mathcal{A}^{\mathfrak{D}}\right)' &= 2\mathcal{H}\,,\label{eqn:pfdebraidA}\\
\left(\ln\mathcal{B}^{\mathfrak{D}}\right)' &= \left[\ln\left(\mathcal{F} + 3\mathcal{H}^2\right)\right]' + 2\mathcal{H}\,,\label{eqn:pfdebraidB}\\
\mathcal{C}^{\mathfrak{D}} &= 0\,,\label{eqn:pfdebraidC}\\
\left(\ln\mathcal{V}^{\mathfrak{D}}\right)' &= \left(\ln\mathcal{I}_2\right)' + 4\mathcal{H}\,,\label{eqn:pfdebraidV}\\
{\alpha^{\mathfrak{D}}}' &= \mathcal{I}_1\mathcal{K}\,.\label{eqn:pfdebraidalpha}
\end{align}
\end{subequations}
From the last line~\eqref{eqn:pfdebraidalpha} we see that the condition \(\alpha^{\mathfrak{D}}(\phi_0) = \alpha^{\mathfrak{D}}_0\), such that
\begin{equation}
\alpha^{\mathfrak{D}}(\phi) = \alpha^{\mathfrak{D}}_0 + \int_{\phi_0}^{\phi}\mathcal{I}_1(\tilde{\phi})\mathcal{K}(\tilde{\phi})\dd\tilde{\phi}\,,
\end{equation}
now uniquely fixes \(\alpha^{\mathfrak{D}}\) independently of the original frame, since it is expressed fully in terms of invariants. Note that this also fixes the remaining parameter functions \(\mathcal{A}^{\mathfrak{D}}, \mathcal{B}^{\mathfrak{D}}, \mathcal{V}^{\mathfrak{D}}\), since they can be expressed in terms of invariants and \(\alpha^{\mathfrak{D}}\) through the definitions~\eqref{eqn:cinv12} and~\eqref{eqn:cinvfh}, and thus take the form
\begin{equation}
\mathcal{A}^{\mathfrak{D}} = \frac{e^{2\alpha^{\mathfrak{D}}}}{\mathcal{I}_1}\,, \quad
\mathcal{B}^{\mathfrak{D}} = 2\frac{e^{2\alpha^{\mathfrak{D}}}}{\mathcal{I}_1}(\mathcal{F} + 3\mathcal{H}^2)\,, \quad
\mathcal{V}^{\mathfrak{D}} = \frac{e^{4\alpha^{\mathfrak{D}}}\mathcal{I}_2}{\mathcal{I}_1^2}\,.
\end{equation}
Finally, it also fixes the frame transition function through \(\gamma^{\mathfrak{D}} = \alpha - \alpha^{\mathfrak{D}}\). Hence, this condition uniquely fixes the debraiding frame and only leaves the freedom to redefine the scalar field. One easily checks that this definition of the debraiding frame is now indeed independent of the original frame, since
\begin{equation}
\theta^{\mathfrak{D}\,a} = e^{\gamma^{\mathfrak{D}}(\phi)}\theta^a = e^{\alpha(\phi) - \alpha^{\mathfrak{D}}(\phi)}\theta^a = e^{-\alpha^{\mathfrak{D}}(\phi)}\theta^{\mathfrak{J}\,a}\,,
\end{equation}
and both \(\alpha^{\mathfrak{D}}\) and the Jordan frame tetrad \(\theta^{\mathfrak{J}\,a}\) are invariants.

We are now in the position to express the action and field equations in the debraiding frame. We start with the action, which now takes the form
\begin{multline}\label{eqn:dfaction}
S^{\mathfrak{D}}\left[\theta^{\mathfrak{D}\,a}, \tp{\omega}^a{}_b, \phi, \chi^I\right] = \frac{1}{2\kappa^2}\int_M\left\{-\frac{e^{2\alpha^{\mathfrak{D}}(\phi)}}{\mathcal{I}_1(\phi)}T^{\mathfrak{D}} + 4\frac{e^{2\alpha^{\mathfrak{D}}(\phi)}}{\mathcal{I}_1(\phi)}[\mathcal{F}(\phi) + 3\mathcal{H}^2(\phi)]X^{\mathfrak{D}} - 2\kappa^2\frac{e^{4\alpha^{\mathfrak{D}}(\phi)}\mathcal{I}_2(\phi)}{\mathcal{I}_1^2(\phi)}\right\}\theta^{\mathfrak{D}}\dd^4x\\
+ S_m^{\mathfrak{J}}\left[e^{\alpha^{\mathfrak{D}}(\phi)}\theta^{\mathfrak{D}\,a}, \chi^I\right]\,,
\end{multline}
and hence does not contain the term \(Y\). We remark that also in this frame one still has the freedom to redefine the scalar field, as it is also the case in the Jordan and Einstein frames we discussed before. We then come to the symmetric part~\eqref{eqn:clafeqtets} of the tetrad field equations, which reads
\begin{equation}\label{eqn:dfeqtets}
2\mathcal{H}S^{\mathfrak{D}}_{(\mu\nu)}{}^{\rho}\phi_{,\rho} + \lc{R}^{\mathfrak{D}}_{\mu\nu} - \frac{1}{2}\lc{R}^{\mathfrak{D}}g^{\mathfrak{D}}_{\mu\nu} + (\mathcal{F} + 3\mathcal{H}^2)\left(\phi_{,\rho}\phi_{,\sigma}g^{\mathfrak{D}\,\rho\sigma}g^{\mathfrak{D}}_{\mu\nu} - 2\phi_{,\mu}\phi_{,\nu}\right) + \kappa^2\frac{e^{2\alpha^{\mathfrak{D}}}\mathcal{I}_2}{\mathcal{I}_1}g^{\mathfrak{D}}_{\mu\nu} = \frac{\kappa^2\mathcal{I}_1}{e^{2\alpha^{\mathfrak{D}}}}\Theta^{\mathfrak{D}}_{\mu\nu}\,.
\end{equation}
The antisymmetric part~\eqref{eqn:clafeqcon}, which is identical to the connection field equations, becomes
\begin{equation}\label{eqn:dfeqcon}
\mathcal{H}T^{\mathfrak{D}\,\rho}{}_{[\mu\nu}\phi_{,\rho]} = 0\,.
\end{equation}
Finally, the scalar field equation~\eqref{eqn:clafeqscal} is given by
\begin{equation}\label{eqn:dfeqscal}
\mathcal{H}T^{\mathfrak{D}} - 2(\mathcal{F} + 3\mathcal{H}^2)\lc{\square}^{\mathfrak{D}}\phi - \left(\mathcal{F}' + 2\mathcal{F}\mathcal{H} + 6\mathcal{H}^3 + 6\mathcal{H}\mathcal{H}'\right)g^{\mathfrak{D}\,\mu\nu}\phi_{,\mu}\phi_{,\nu} + \kappa^2\frac{e^{2\alpha^{\mathfrak{D}}}}{\mathcal{I}_1}(4\mathcal{I}_2\mathcal{H} + \mathcal{I}_2') = \frac{\kappa^2\mathcal{I}_1^2\mathcal{K}}{e^{2\alpha^{\mathfrak{D}}}}\Theta^{\mathfrak{D}}\,.
\end{equation}
We see that now indeed the tetrad and scalar field equations are debraided, i.e., the former contains only second derivatives of the tetrad, while the latter contains only second derivatives of the scalar field.

We conclude our discussion of the debraiding frame with a final remark. One may argue that this frame could more rightfully be called the Einstein frame, since the debraiding essentially turns the scalar field into another source term for the tetrad instead of interrelating their dynamics. One could equally well argue that there is no Einstein frame, since even in the debraiding frame the scalar field is non-minimally coupled to torsion through the term \(\mathcal{A}^{\mathfrak{D}}(\phi)T\) in the action. However, we will not enter this discussion here, as it is merely a question of nomenclature.

This concludes our discussion of scalar-torsion theories of gravity with a single field coupled to the tetrad and the spin connection. It is natural to ask whether the results we obtained also apply to theories with multiple scalar fields. This question will be explored in the following section.

\section{Generalization to multiple scalar fields}\label{sec:multi}
In the previous sections we have considered a single scalar field in addition to the tetrad and spin connection as the dynamical variables of the gravitational interaction. We now generalize our statements and results to multiple scalar fields. This will be done in several steps. We define the generalized action in section~\ref{ssec:maction}, and derive the corresponding field equations in section~\ref{ssec:mfeqs}. Conformal transformations are discussed in section~\ref{ssec:mconformal}. From these we finally derive invariants in section~\ref{ssec:minv} and discuss particular conformal frames in section~\ref{ssec:mframes}.

\subsection{Action}\label{ssec:maction}
Instead of a single scalar field \(\phi\) we now consider a scalar field multiplet \(\boldsymbol{\phi} = (\phi^A, A = 1, \ldots, N)\) of \(N\) scalar fields. This imposes two changes to the class of scalar-torsion theories defined by the action~\eqref{eqn:classactiong} and~\eqref{eqn:classactionm}. First, instead of the single kinetic and derivative coupling terms \(X\) and \(Y\) one may now form the terms
\begin{equation}
X^{AB} = -\frac{1}{2}g^{\mu\nu}\phi^A_{,\mu}\phi^B_{,\nu}\,, \quad Y^A = T_{\mu}{}^{\mu\nu}\phi^A_{,\nu}\,,
\end{equation}
making use of all scalar fields. Note that \(X^{AB}\) is symmetric, \(X^{[AB]} = 0\). Second, the free parameter functions on the action may now depend on all scalar fields. Hence, we generalize the action~\eqref{eqn:classactiong} such that it reads
\begin{equation}\label{eqn:multiclassactiong}
S_g\left[\theta^a, \tp{\omega}^a{}_b, \phi^A\right] = \frac{1}{2\kappa^2}\int_M\left[-\mathcal{A}(\boldsymbol{\phi})T + 2\mathcal{B}_{AB}(\boldsymbol{\phi})X^{AB} + 2\mathcal{C}_A(\boldsymbol{\phi})Y^A - 2\kappa^2\mathcal{V}(\boldsymbol{\phi})\right]\theta\dd^4x\,.
\end{equation}
We remark that now also the functions \(\mathcal{B}_{AB}\) and \(\mathcal{C}_A\) carry scalar field indices, which are contracted with the corresponding indices of \(X^{AB}\) and \(Y^A\). Note that \(\mathcal{B}_{AB}\) must be symmetric, \(\mathcal{B}_{[AB]} = 0\), since any antisymmetric contribution would cancel in the contraction with the symmetric term \(X^{AB}\). Also in the matter action~\eqref{eqn:classactionm} the free function \(\alpha\), which determines the conformally related tetrad coupled to matter, now depends on all scalar fields,
\begin{equation}\label{eqn:multiclassactionm}
S_m[\theta^a, \phi^A, \chi^I] = S_m^{\mathfrak{J}}\left[e^{\alpha(\boldsymbol{\phi})}\theta^a, \chi^I\right]\,.
\end{equation}
The particular form of the matter action now imposes a relation between the sources \(\vartheta_A\) in the scalar field equations, which are obtained from the variation
\begin{equation}\label{eqn:multicmatactvar}
\delta S_m[\theta^a, \phi^A, \chi^I] = \int_M\left(\Theta_a{}^{\mu}\delta\theta^a{}_{\mu} + \vartheta_A\delta\phi^A + \varpi_I\delta\chi^I\right)\theta\dd^4x\,,
\end{equation}
and the energy-momentum tensor \(\Theta_a{}^{\mu}\), which generalizes the relation~\eqref{eqn:classenmomtens} and reads
\begin{equation}\label{eqn:mclassenmomtens}
\vartheta_A = \alpha_{,A}\theta^a{}_{\mu}\Theta_a{}^{\mu}\,.
\end{equation}
This relation will be used during the remainder of this section.

We now also see why we favored the form~\eqref{eqn:classactiong} over the equivalent form~\eqref{eqn:classactionb}. A similar generalization of the latter to multiple scalar fields would yield an action of the form
\begin{equation}\label{eqn:multiclassactionb}
S_g\left[\theta^a, \tp{\omega}^a{}_b, \phi^A\right] = \frac{1}{2\kappa^2}\int_M\left[-\mathcal{A}(\boldsymbol{\phi})T + 2\mathcal{B}_{AB}(\boldsymbol{\phi})X^{AB} - \tilde{\mathcal{C}}(\boldsymbol{\phi})B - 2\kappa^2\mathcal{V}(\boldsymbol{\phi})\right]\theta\dd^4x\,,
\end{equation}
which is equivalent to the action~\eqref{eqn:multiclassactiong} (up to a boundary term) only if \(\mathcal{C}_A = \tilde{\mathcal{C}}_{,A}\), where we use the comma notation to indicate derivatives with respect to scalar fields \(\phi^A\). This imposes an additional restriction on the parameter functions \(\mathcal{C}_A\), and in particular implies \(\mathcal{C}_{[A,B]} = 0\). Here we will not make this restriction, and work with the action~\eqref{eqn:multiclassactiong} with arbitrary parameter functions \(\mathcal{C}_A\).

\subsection{Field equations}\label{ssec:mfeqs}
We can then proceed with the field equations for the multi-scalar-torsion theories. As we did in the single field case in section~\ref{sec:feqs}, we omit their derivation here, since the action~\eqref{eqn:multiclassactiong} is a special case of the more general multi-scalar-torsion generalization of the \(L(T, X, Y, \phi)\) class of theories~\cite{Hohmann:2018dqh}, where the Lagrangian is given by
\begin{equation}
L = \frac{1}{2\kappa^2}\left[-\mathcal{A}(\boldsymbol{\phi})T + 2\mathcal{B}_{AB}(\boldsymbol{\phi})X^{AB} + 2\mathcal{C}_A(\boldsymbol{\phi})Y^A\right] - \mathcal{V}(\boldsymbol{\phi})\,.
\end{equation}
Hence, we can make use of the field equations derived for this more general theory, together with the variational derivatives
\begin{equation}
L_T = -\frac{\mathcal{A}}{2\kappa^2}\,, \quad L_{X^{AB}} = \frac{\mathcal{B}_{AB}}{\kappa^2}\,, \quad L_{Y^A} = \frac{\mathcal{C}_A}{\kappa^2}\,, \quad L_{\phi^A} = \frac{1}{2\kappa^2}\left[-\mathcal{A}_{,A}T + 2\mathcal{B}_{BC,A}X^{BC} + 2\mathcal{C}_{B,A}Y^B\right] - \mathcal{V}_{,A}\,.
\end{equation}
Here we restrict ourselves to displaying the final form of the field equation as given in section~\ref{sec:feqs}. For the symmetric part~\eqref{eqn:clafeqtets} we find
\begin{multline}\label{eqn:multiclafeqtets}
\left(\mathcal{A}_{,A} + \mathcal{C}_A\right)S_{(\mu\nu)}{}^{\rho}\phi^A_{,\rho} + \mathcal{A}\left(\lc{R}_{\mu\nu} - \frac{1}{2}\lc{R}g_{\mu\nu}\right) + \left(\frac{1}{2}\mathcal{B}_{AB} - \mathcal{C}_{(A,B)}\right)\phi^A_{,\rho}\phi^B_{,\sigma}g^{\rho\sigma}g_{\mu\nu}\\
- \left(\mathcal{B}_{AB} - \mathcal{C}_{(A,B)}\right)\phi^A_{,\mu}\phi^B_{,\nu} + \mathcal{C}_A\left(\lc{\nabla}_{\mu}\lc{\nabla}_{\nu}\phi^A - \lc{\square}\phi^Ag_{\mu\nu}\right) + \kappa^2\mathcal{V}g_{\mu\nu} = \kappa^2\Theta_{\mu\nu}\,.
\end{multline}
while the antisymmetric part~\eqref{eqn:clafeqcon} reads
\begin{equation}\label{eqn:multiclafeqcon}
3(\mathcal{A}_{,A} + \mathcal{C}_A)T^{\rho}{}_{[\mu\nu}\phi^A_{,\rho]} + 2\mathcal{C}_{[A,B]}\phi^A_{,\mu}\phi^B_{,\nu} = 0\,,
\end{equation}
and the scalar field equations~\eqref{eqn:clafeqscal} are given by
\begin{equation}\label{eqn:multiclafeqscal}
\frac{1}{2}\mathcal{A}_{,A}T - \mathcal{B}_{AB}\lc{\square}\phi^B - \left(\mathcal{B}_{AB,C} - \frac{1}{2}\mathcal{B}_{BC,A}\right)g^{\mu\nu}\phi^B_{,\mu}\phi^C_{,\nu} + \mathcal{C}_A\lc{\nabla}_{\mu}T_{\nu}{}^{\nu\mu} + 2\mathcal{C}_{[A,B]}T_{\mu}{}^{\mu\nu}\phi^B_{,\nu} + \kappa^2\mathcal{V}_{,A} = \kappa^2\alpha_{,A}\Theta\,.
\end{equation}
Note the appearance of a few additional terms containing \(\mathcal{C}_{[A,B]}\), which do not appear in the single field case detailed in section~\ref{sec:feqs}, since they vanish due to the antisymmetrization brackets, and which would also vanish if we used the action~\eqref{eqn:multiclassactionb}. Finally, we may also perform a debraiding of the scalar field equations, i.e., remove the second order derivatives of the tetrad by adding a suitable multiple of the trace
\begin{equation}
-2\left(\mathcal{A}_{,A} + \mathcal{C}_A\right)T_{\mu}{}^{\mu\nu}\phi^A_{,\nu} - \mathcal{A}\lc{R} + \left(\mathcal{B}_{AB} - 3\mathcal{C}_{A,B}\right)g^{\mu\nu}\phi^A_{,\mu}\phi^B_{,\nu} - 3\mathcal{C}_A\lc{\square}\phi^A + 4\kappa^2\mathcal{V} = \kappa^2\Theta\,.
\end{equation}
The resulting field equations then take the form
\begin{multline}\label{eqn:multiclafeqdeb}
\mathcal{A}(\mathcal{A}_{,A} + \mathcal{C}_A)T + \left[4\mathcal{A}\mathcal{C}_{[A,B]} - 2\mathcal{C}_A(\mathcal{A}_{,B} + \mathcal{C}_B)\right]T_{\mu}{}^{\mu\nu}\phi^B_{,\nu} - \left(2\mathcal{A}\mathcal{B}_{AB} + 3\mathcal{C}_A\mathcal{C}_B\right)\lc{\square}\phi^B\\
+ (\mathcal{C}_A\mathcal{B}_{BC} - 2\mathcal{A}\mathcal{B}_{AB,C} + \mathcal{A}\mathcal{B}_{BC,A} - 3\mathcal{C}_A\mathcal{C}_{B,C})g^{\mu\nu}\phi^B_{,\mu}\phi^C_{,\nu} + 2\kappa^2(\mathcal{A}\mathcal{V}_{,A} + 2\mathcal{C}_A\mathcal{V}) = \kappa^2(2\mathcal{A}\alpha_A + \mathcal{C}_A)\Theta\,.
\end{multline}
One may pose the question whether also the second derivatives of the scalar field can be removed from the tetrad field equations~\eqref{eqn:multiclafeqtets} in a suitable frame; we will see in section~\ref{ssec:mconformal} that this is not always possible.

\subsection{Conformal transformations}\label{ssec:mconformal}
Turning our attention to conformal transformations, we see that also the action~\eqref{eqn:multiclassactiong} and~\eqref{eqn:multiclassactionm} retains its form under conformal transformations and scalar field redefinitions given by
\begin{equation}\label{eqn:multiconftrans}
\bar{\theta}^a{}_{\mu} = e^{\gamma(\boldsymbol{\phi})}\theta^a{}_{\mu}\,, \quad \bar{e}_a{}^{\mu} = e^{-\gamma(\boldsymbol{\phi})}e_a{}^{\mu}\,, \quad \bar{\phi}^A = f^A(\boldsymbol{\phi})\,,
\end{equation}
in the same sense as explained in detail in section~\ref{sec:conformal}. In the following we will also collectively write \(\bar{\boldsymbol{\phi}} = \boldsymbol{f}(\boldsymbol{\phi})\) for the scalar field redefinition. Proceeding in analogy to the calculation~\eqref{eqn:classacttransg} and~\eqref{eqn:classacttransm} and comparing the transformed action to its original form, we find that the functions parametrizing the action must transform as
\begin{subequations}\label{eqn:multipftrans}
\begin{align}
\mathcal{A} &= e^{2\gamma}\bar{\mathcal{A}}\,,\label{eqn:multipftransA}\\
\mathcal{B}_{AB} &= e^{2\gamma}\left(\bar{\mathcal{B}}_{CD}\frac{\partial\bar{\phi}^C}{\partial\phi^A}\frac{\partial\bar{\phi}^D}{\partial\phi^B} - 6\bar{\mathcal{A}}\gamma_{,A}\gamma_{,B} + 6\bar{\mathcal{C}}_C\frac{\partial\bar{\phi}^C}{\partial\phi^{(A}}\gamma_{,B)}\right)\,,\label{eqn:multipftransB}\\
\mathcal{C}_A &= e^{2\gamma}\left(\bar{\mathcal{C}}_B\frac{\partial\bar{\phi}^B}{\partial\phi^A} - 2\bar{\mathcal{A}}\gamma_{,A}\right)\,,\label{eqn:multipftransC}\\
\mathcal{V} &= e^{4\gamma}\bar{\mathcal{V}}\,,\label{eqn:multipftransV}\\
\alpha &= \bar{\alpha} + \gamma\,.\label{eqn:multipftransalpha}
\end{align}
\end{subequations}
This transformation behavior generalizes the relations~\eqref{eqn:pftrans}. Note that instead of the derivative \(f'\) we now find the Jacobian of the function~\(\boldsymbol{f}\).

Finally, we remark that also in the case of multiple scalar fields the corresponding relation~\eqref{eqn:mclassenmomtens} between the source terms in the field equations is preserved under their conformal transformation, which is given by
\begin{equation}\label{eqn:multimattermtrans}
\Theta_{\mu\nu} = e^{2\gamma}\bar{\Theta}_{\mu\nu}\,, \quad
\Theta = e^{4\gamma}\bar{\Theta}\,, \quad
\vartheta_A = e^{4\gamma}\left(\gamma_{,A}\bar{\Theta} + \frac{\partial\bar{\phi}^B}{\partial\phi^A}\bar{\vartheta}_B\right)\,,
\end{equation}
which generalizes the transformation rule~\eqref{eqn:mattermtrans}.

\subsection{Invariant quantities}\label{ssec:minv}
The form of the transformations~\eqref{eqn:multipftrans} motivates the definition of a number of quantities which are invariant under conformal transformations and either invariant or covariant under redefinitions of the scalar fields, proceeding in full analogy to the quantities defined in section~\ref{sec:invariant}. First note that the transformation behavior~\eqref{eqn:multipftransA}, \eqref{eqn:multipftransV} and~\eqref{eqn:multipftransalpha} of the functions \(\mathcal{A}, \mathcal{V}, \alpha\) agrees with the single field case given by the relations~\eqref{eqn:pftransA}, \eqref{eqn:pftransV} and~\eqref{eqn:pftransalpha}. Hence, the quantities \(\mathcal{I}_1\) and \(\mathcal{I}_2\) retain their invariant character, and we keep their definitions~\eqref{eqn:cinv12}. For the remaining quantities, which carry scalar field indices in the case of multiple scalar fields, we must adapt their definitions. For \(\mathcal{F}\) and \(\mathcal{H}\) we extend the definitions~\eqref{eqn:cinvfh} to
\begin{equation}\label{eqn:multicinvfh}
\mathcal{F}_{AB} = \frac{2\mathcal{A}\mathcal{B}_{AB} - 6\mathcal{A}_{,(A}\mathcal{C}_{B)} - 3\mathcal{A}_{,A}\mathcal{A}_{,B}}{4\mathcal{A}^2}\,, \quad
\mathcal{H}_A = \frac{\mathcal{C}_A + \mathcal{A}_{,A}}{2\mathcal{A}}\,,
\end{equation}
while the definitions~\eqref{eqn:cinvgk} of \(\mathcal{G}\) and \(\mathcal{K}\) generalize to
\begin{equation}\label{eqn:multicinvgk}
\mathcal{G}_{AB} = \frac{\mathcal{B}_{AB} - 6\alpha_{,(A}\mathcal{C}_{B)} - 6\alpha_{,A}\alpha_{,B}\mathcal{A}}{2e^{2\alpha}}\,, \quad
\mathcal{K}_A = \frac{\mathcal{C}_A + 2\alpha_{,A}\mathcal{A}}{2e^{2\alpha}}\,.
\end{equation}
By comparison with the transformations~\eqref{eqn:multipftrans} we then see that these quantities are invariant under conformal transformations, but transform covariantly under scalar field redefinitions,
\begin{equation}\label{eqn:multifhgktrans}
\bar{\mathcal{F}}_{AB} = \frac{\partial\phi^C}{\partial\bar{\phi}^A}\frac{\partial\phi^D}{\partial\bar{\phi}^B}\mathcal{F}_{CD}\,, \quad
\bar{\mathcal{H}}_A = \frac{\partial\phi^B}{\partial\bar{\phi}^A}\mathcal{H}_B\,, \quad
\bar{\mathcal{G}}_{AB} = \frac{\partial\phi^C}{\partial\bar{\phi}^A}\frac{\partial\phi^D}{\partial\bar{\phi}^B}\mathcal{G}_{CD}\,, \quad
\bar{\mathcal{K}}_A = \frac{\partial\phi^B}{\partial\bar{\phi}^A}\mathcal{K}_B\,,
\end{equation}
where we again encounter the inverse Jacobian of the function \(\boldsymbol{f}\). It is worth noting that this transformation behavior has the same form as that of tensor fields on a manifold, whose points are the values of the multiplet of scalar fields, under general coordinate transformations. However, we will not pursue this interpretation here, as it would exceed the scope of this article. We also remark that the quantities~\eqref{eqn:multicinvfh} and~\eqref{eqn:multicinvgk} are related to each other by
\begin{equation}\label{eqn:multifhgk}
\mathcal{F}_{AB} = \mathcal{I}_1\mathcal{G}_{AB} + 3\frac{\mathcal{I}_1'}{\mathcal{I}_1}\left(\mathcal{I}_1\mathcal{K} - \frac{\mathcal{I}_1'}{4\mathcal{I}_1}\right)\,, \quad
\mathcal{H}_A = \mathcal{I}_1\mathcal{K}_A - \frac{\mathcal{I}_{1,A}}{2\mathcal{I}_1}\,,
\end{equation}
which generalizes the similar relations~\eqref{eqn:fhgk}.

\subsection{Conformal frames}\label{ssec:mframes}
We finally also generalize the particular conformal frames discussed in section~\ref{sec:frames} to multiple scalar fields. For the Jordan frame shown in section~\ref{ssec:jordan} this is straightforward. Starting from the conformal transformation~\eqref{eqn:jframedef} we find that the relations~\eqref{eqn:pfjordan} generalize to
\begin{equation}\label{eqn:mpfjordan}
\mathcal{A}^{\mathfrak{J}} = \frac{1}{\mathcal{I}_1}\,, \quad
\mathcal{B}_{AB}^{\mathfrak{J}} = 2\mathcal{G}_{AB}\,, \quad
\mathcal{C}_A^{\mathfrak{J}} = 2\mathcal{K}_A\,, \quad
\mathcal{V}^{\mathfrak{J}} = \frac{\mathcal{I}_2}{\mathcal{I}_1^2}\,, \quad
\alpha^{\mathfrak{J}} = 0\,.
\end{equation}
Also the Einstein frame detailed in section~\ref{ssec:einstein} immediately generalizes. From the transformation~\eqref{eqn:jframedef} we find the parameter functions
\begin{equation}\label{eqn:mpfeinstein}
\mathcal{A}^{\mathfrak{E}} = 1\,, \quad
\mathcal{B}_{AB}^{\mathfrak{E}} = 2\mathcal{F}_{AB}\,, \quad
\mathcal{C}_A^{\mathfrak{E}} = 2\mathcal{H}_A\,, \quad
\mathcal{V}^{\mathfrak{E}} = \mathcal{I}_2\,, \quad
\alpha^{\mathfrak{E}} = \frac{1}{2}\ln\mathcal{I}_1\,.
\end{equation}
Proceeding in analogy to section~\ref{sec:frames}, one may now express the action shown in section~\ref{ssec:maction} and field equations shown in section~\ref{ssec:mfeqs} in these conformal frames. We will not explicitly display the result here, as it is essentially the same as in the single field case and easy to derive.

The situation is qualitatively different for the debraiding frame introduced in section~\ref{ssec:debraiding}. One can see from the symmetric tetrad field equation~\eqref{eqn:multiclafeqtets} that the second order derivatives of the scalar fields can be removed from the tetrad field equations in a particular ``debraiding'' frame \(\mathfrak{D}\) only by imposing \(\mathcal{C}_A^{\mathfrak{D}} = 0\). By comparison to the transformations~\eqref{eqn:multipftransA} and~\eqref{eqn:multipftransC} we then find the condition
\begin{equation}
\gamma^{\mathfrak{D}}_{,A}(\phi) = -\frac{\mathcal{C}_A(\phi)}{2\mathcal{A}(\phi)}\,,
\end{equation}
which can be satisfied only if there exists some function \(\tilde{\mathcal{H}}\) such that \(\mathcal{H}_A = \tilde{\mathcal{H}}_{,A}\).

This concludes our general discussion of scalar-torsion and multi-scalar-torsion theories of gravity. In order to show the applicability of our results and relate them to other works, we will provide a few examples in the following section.

\section{Examples}\label{sec:examples}
We finally connect our results to a number of example theories. Note that some authors use different sign conventions, in particular for the signature of the metric tensor; however, these can simply be absorbed into a suitable redefinition of the parameter functions in the action. Here we discuss teleparallel dark energy and its generalizations in section~\ref{ssec:darkenergy}, include a non-minimal coupling to the boundary term in section~\ref{ssec:boundcoup} and come to the multi-scalar-torsion equivalent of \(F(T, B)\) gravity theories in section~\ref{ssec:ftb}. Finally, we show how our results reduce to a number of well-known results in multi-scalar-tensor gravity in section~\ref{ssec:stg}.

\subsection{Teleparallel dark energy and its generalizations}\label{ssec:darkenergy}
The first example we show is the classical teleparallel dark energy model~\cite{Geng:2011aj}. Its action can be written in the form
\begin{equation}
S_g = \int_M\left[-\frac{T}{2\kappa^2} + \frac{1}{2}\left(g^{\mu\nu}\phi_{,\mu}\phi_{,\nu} - \xi\phi^2T\right) - V(\phi)\right]\theta\dd^4x\,,
\end{equation}
with coupling constant \(\xi\) and potential \(V\). By comparison with the general form~\eqref{eqn:classactiong} we find the parameter functions
\begin{equation}
\mathcal{A} = 1 + 2\kappa^2\xi\phi^2\,, \quad
\mathcal{B} = -\kappa^2\,, \quad
\mathcal{C} = 0\,, \quad
\mathcal{V} = V\,.
\end{equation}
One usually considers this model to be given in the Jordan frame, such that \(\alpha = 0\). Various generalizations of this model has been considered:
\begin{enumerate}
\item
Interacting dark energy~\cite{Otalora:2013tba}:
\begin{equation}
S_g = \int_M\left[-\frac{T}{2\kappa^2} + \frac{1}{2}\left(g^{\mu\nu}\phi_{,\mu}\phi_{,\nu} - \xi F(\phi)T\right) - V(\phi)\right]\theta\dd^4x\,,
\end{equation}
where the function \(\mathcal{A}\) is replaced by \(\mathcal{A} = 1 + 2\kappa^2\xi F(\phi)\).

\item
Brans-Dicke type action with a general coupling to torsion~\cite{Izumi:2013dca}:
\begin{equation}
S_g = \int_M\left[-\frac{F(\phi)}{2\kappa^2}T - \omega g^{\mu\nu}\phi_{,\mu}\phi_{,\nu} - V(\phi)\right]\theta\dd^4x\,,
\end{equation}
where \(\mathcal{A} = F(\phi)\) and \(\mathcal{B} = 2\kappa^2\omega\).

\item
Brans-Dicke type action with a dynamical kinetic term~\cite{Chen:2014qsa}:
\begin{equation}
S_g = \int_M\left[-\frac{\phi}{2\kappa^2}T - \frac{\omega(\phi)}{\phi}g^{\mu\nu}\phi_{,\mu}\phi_{,\nu} - V(\phi)\right]\theta\dd^4x\,,
\end{equation}
where \(\mathcal{A} = \phi\) and \(\mathcal{B} = 2\kappa^2\omega(\phi)/\phi\).
\end{enumerate}
Note that all these models satisfy \(\mathcal{K} = 0\), and so are considered minimally coupled according to our convention, despite their non-minimal coupling between the scalar field and the torsion scalar. This is due to the fact that this type of coupling does not introduce the trace \(\Theta\) of the energy-momentum tensor as a source into the debraided scalar field equation~\eqref{eqn:clafeqdeb}.

\subsection{Non-minimal coupling to the boundary term}\label{ssec:boundcoup}
In addition to the torsion scalar, as in the original teleparallel dark energy model~\cite{Geng:2011aj} discussed above, one may also include a similar type of coupling to the boundary term \(B = \lc{R} + T = 2\lc{\nabla}_{\mu}T^{\mu\nu}{}_{\nu}\). The corresponding action functional reads~\cite{Bahamonde:2015hza}
\begin{equation}
S_g = \int_M\left[-\frac{T}{2\kappa^2} + \frac{1}{2}\left(g^{\mu\nu}\phi_{,\mu}\phi_{,\nu} - \xi\phi^2T - \chi\phi^2B\right) - V(\phi)\right]\theta\dd^4x
\end{equation}
with constants \(\xi, \chi\) and a general potential \(V\). We see that this action is of the form~\eqref{eqn:classactionb}, with parameter functions given by
\begin{equation}
\mathcal{A} = 1 + 2\kappa^2\xi\phi^2\,, \quad
\mathcal{B} = -\kappa^2\,, \quad
\tilde{\mathcal{C}} = 2\kappa^2\chi\phi^2\,, \quad
\mathcal{V} = V\,.
\end{equation}
It follows that the action may be brought to the form~\eqref{eqn:classactiong} by integration by parts, where the remaining parameter functions becomes
\begin{equation}
\mathcal{C} = \tilde{\mathcal{C}}' = 4\kappa^2\chi\phi\,.
\end{equation}
Note that for \(\xi + \chi = 0\) the action reduces to scalar-tensor gravity with a non-minimally coupled scalar field, while for \(\chi = 0\) one obtains the teleparallel dark energy model~\cite{Geng:2011aj} shown in section~\ref{ssec:darkenergy}. Also in this case one usually considers \(\alpha = 0\). We further remark that also more general models with multiple scalar fields coupled to the boundary term are considered, which may similarly be written in the form~\eqref{eqn:multiclassactionb}~\cite{Bahamonde:2018miw}.

\subsection{Scalar-torsion equivalent of $F(T, B)$ gravity}\label{ssec:ftb}
A more general action involving the boundary term \(B\) is given by \(F(T, B)\) gravity and reads~\cite{Wright:2016ayu}
\begin{equation}
S_g = \frac{1}{2\kappa^2}\int_MF(T, B)\theta\dd^4x\,.
\end{equation}
In order to bring it to the form~\eqref{eqn:classactionb} one introduces two auxiliary scalar fields \(\phi_{1,2}\), and replaces the arguments of \(F\) with these fields. Enforcing \(\phi_1 = T\) and \(\phi_2 = B\) via Lagrange multipliers and eliminating the Lagrange multipliers from the action one obtains
\begin{equation}
S_g = \frac{1}{2\kappa^2}\int_M\left[F^{(1,0)}(\boldsymbol{\phi})T + F^{(0,1)}(\boldsymbol{\phi})B + F(\boldsymbol{\phi}) - \phi_1F^{(1,0)}(\boldsymbol{\phi}) - \phi_2F^{(0,1)}(\boldsymbol{\phi})\right]\theta\dd^4x\,.
\end{equation}
Comparison with the action~\eqref{eqn:classactionb} yields the parameter functions
\begin{equation}
\mathcal{A} = -F^{(1,0)}\,, \quad
\mathcal{B} = 0\,, \quad
\tilde{\mathcal{C}} = -F^{(0,1)}\,, \quad
\mathcal{V} = \frac{1}{2\kappa^2}\left(\phi_1F^{(1,0)} + \phi_2F^{(0,1)} - F\right)\,.
\end{equation}
Again we can integrate by parts to bring the action to the form~\eqref{eqn:classactiong}, and finally obtain
\begin{equation}
\mathcal{C}_1 = -F^{(1,1)}\,, \quad
\mathcal{C}_2 = -F^{(0,2)}\,.
\end{equation}
We also remark that in the case that \(F\) does not depend on its second argument the scalar field \(\phi_2\) drops out, and the action reduces to the scalar-torsion equivalent of \(F(T)\) gravity~\cite{Izumi:2013dca}.

\subsection{(Multi-)scalar-tensor gravity}\label{ssec:stg}
We finally discuss a special case for the function \(\mathcal{C}\), which is given by the relation \(\mathcal{C} = -\mathcal{A}'\), and which can invariantly be formulated as \(\mathcal{H} = 0\). In this case the terms containing \(T\) and \(Y\) in the action~\eqref{eqn:classactiong} can be combined,
\begin{equation}
-\mathcal{A}T - 2\mathcal{A}'Y = -\mathcal{A}T - 2\partial_{\mu}\mathcal{A}T_{\nu}{}^{\nu\mu} = \mathcal{A}\left(2\lc{\nabla}_{\mu}T_{\nu}{}^{\nu\mu} - T\right) - 2\lc{\nabla}_{\mu}\left(\mathcal{A}T_{\nu}{}^{\nu\mu}\right) = \mathcal{A}\lc{R} - 2\lc{\nabla}_{\mu}\left(\mathcal{A}T_{\nu}{}^{\nu\mu}\right)\,.
\end{equation}
It follows that up to a boundary term, which we neglect here, the gravitational part of the action reduces to the well-known scalar-tensor gravity action~\cite{Flanagan:2004bz}
\begin{equation}\label{eqn:stgactiong}
S_g\left[\theta^a, \tp{\omega}^a{}_b, \phi\right] = \frac{1}{2\kappa^2}\int_M\left[\mathcal{A}(\phi)\lc{R} + 2\mathcal{B}(\phi)X - 2\kappa^2\mathcal{V}(\phi)\right]\theta\dd^4x\,.
\end{equation}
This becomes apparent also at the level of the field equations. In the symmetric field equation~\eqref{eqn:clafeqtets} the terms involving the superpotential cancel, and the remaining terms take the usual form
\begin{multline}\label{eqn:stgfeqmet}
\mathcal{A}\left(\lc{R}_{\mu\nu} - \frac{1}{2}\lc{R}g_{\mu\nu}\right) + \left(\frac{1}{2}\mathcal{B} + \mathcal{A}''\right)\phi_{,\rho}\phi_{,\sigma}g^{\rho\sigma}g_{\mu\nu}\\
- (\mathcal{B} + \mathcal{A}'')\phi_{,\mu}\phi_{,\nu} - \mathcal{A}'\left(\lc{\nabla}_{\mu}\lc{\nabla}_{\nu}\phi - \lc{\square}\phi g_{\mu\nu}\right) + \kappa^2\mathcal{V}g_{\mu\nu} = \kappa^2\Theta_{\mu\nu}\,.
\end{multline}
The connection field equations~\eqref{eqn:clafeqcon} are identically satisfied, since the action~\eqref{eqn:stgactiong} is independent of the spin connection. Finally, also the scalar field equation~\eqref{eqn:clafeqscal} reduces to its well-known scalar-tensor form, which reads
\begin{equation}\label{eqn:stgfeqscal}
-\frac{1}{2}\mathcal{A}'\tp{R} - \mathcal{B}\lc{\square}\phi - \frac{1}{2}\mathcal{B}'g^{\mu\nu}\phi_{,\mu}\phi_{,\nu} + \kappa^2\mathcal{V}' = \kappa^2\alpha'\Theta\,.
\end{equation}
We finally remark that in this case also the invariant quantities introduced in section~\ref{sec:invariant} reduce to their scalar-tensor counterparts~\cite{Jarv:2014hma}.

One easily checks that also the multi-scalar-torsion action~\eqref{eqn:multiclassactiong} allows for a similar choice of the parameter functions, which is given by the condition \(\mathcal{C}_A = -\mathcal{A}_{,A}\) and thus generalizes the scalar-tensor condition discussed above. In terms of invariants this condition is expressed as \(\mathcal{H}_A = 0\). In this case the action reduces in a similar fashion as the action~\eqref{eqn:stgactiong} and now becomes
\begin{equation}\label{eqn:mstgactiong}
S_g\left[\theta^a, \tp{\omega}^a{}_b, \phi^A\right] = \frac{1}{2\kappa^2}\int_M\left[\mathcal{A}(\phi)\lc{R} + 2\mathcal{B}_{AB}(\phi)X^{AB} - 2\kappa^2\mathcal{V}(\phi)\right]\theta\dd^4x\,.
\end{equation}
From this one recognizes the action functional of multi-scalar-tensor gravity~\cite{Damour:1992we,Berkin:1993bt}, with metric field equation given by
\begin{multline}\label{eqn:mstgfeqmet}
\mathcal{A}\left(\lc{R}_{\mu\nu} - \frac{1}{2}\lc{R}g_{\mu\nu}\right) + \left(\frac{1}{2}\mathcal{B}_{AB} + \mathcal{A}_{,AB}\right)\phi^A_{,\rho}\phi^B_{,\sigma}g^{\rho\sigma}g_{\mu\nu}\\
- (\mathcal{B}_{AB} + \mathcal{A}_{,AB})\phi^A_{,\mu}\phi^B_{,\nu} - \mathcal{A}_{,A}\left(\lc{\nabla}_{\mu}\lc{\nabla}_{\nu}\phi^A - \lc{\square}\phi^Ag_{\mu\nu}\right) + \kappa^2\mathcal{V}g_{\mu\nu} = \kappa^2\Theta_{\mu\nu}\,,
\end{multline}
while the scalar field equation reduces to
\begin{equation}\label{eqn:mstgfeqscal}
-\frac{1}{2}\mathcal{A}_A\tp{R} - \mathcal{B}_{AB}\lc{\square}\phi^B - \left(\mathcal{B}_{AB,C} - \frac{1}{2}\mathcal{B}_{BC,A}\right)g^{\mu\nu}\phi^B_{,\mu}\phi^C_{,\nu} + \kappa^2\mathcal{V}_{,A} = \kappa^2\alpha_{,A}\Theta\,.
\end{equation}
Finally, one finds that the invariants introduced in section~\ref{sec:invariant} reduce to their multi-scalar-tensor expressions~\cite{Kuusk:2015dda}. We also remark that the invariant \(\mathcal{K}_A\) reduces to the vector of non-minimal coupling defined in~\cite{Hohmann:2016yfd}.

This concludes our discussion of example theories. We have seen that the framework we developed in this article has a wide range of possible future applications, and that it reduces to the known calculations in (multi)-scalar-tensor for a suitably chosen class of actions.

\section{Conclusion}\label{sec:conclusion}
In this article we have discussed a class of teleparallel scalar-torsion theories of gravity defined by five free functions of the scalar field, whose action has a similar structure to that of scalar-tensor gravity. We have studied their field equations and behavior under conformal transformations of the tetrad, as well as redefinitions of the scalar field. In particular, we have shown that such transformations relate different theories of this class, defined by a set of transformed parameter functions, to each other. As one of the main results we have derived a number of functions of the scalar field, which are composed from the free functions in the action, and which are either invariant or transform covariantly under these transformations. Further, we have discussed different conformal frames, and obtained conditions for minimally coupling of the scalar field and for separating the highest order derivatives in the field equations. We also generalized our results to multiple scalar fields.

The framework of invariants we developed generalizes the formerly developed framework of invariants in scalar-tensor and multi-scalar-tensor gravity theories~\cite{Jarv:2014hma,Kuusk:2015dda}. It allows to easily translate the action and field equations of any scalar-torsion theory of gravity, defined in an arbitrary frame, to any other frame, and in particular to the Jordan and Einstein frames known from scalar-tensor gravity. We have also shown that our framework reduces to the (multi-)scalar-tensor framework of invariants in the case that one of the scalar-torsion invariants vanishes. We expect this framework to be of the same use in describing phenomenological aspects of scalar-torsion gravity in a frame independent fashion, as it is also the case for its scalar-tensor counterpart~\cite{Jarv:2015kga,Kuusk:2016rso,Jarv:2016sow}.

As another interesting result we have shown that a naively defined Einstein frame, in which there is no direct coupling between the scalar field and the torsion scalar, does not lead to a ``debraiding'' of the field equations as it is the case in scalar-tensor theories~\cite{Bettoni:2015wta}. Instead, debraiding is achieved in a different class of frames, in which the coefficient of the kinetic coupling term of the scalar field vanishes, and which is defined only up to a free parameter. We also demonstrated that in the case of multiple scalar fields such a frame choice is possible only for a restricted class of actions.

Our results invite for a number of further studies of the class of theories we discussed. From the phenomenological point of view, observational properties such as the post-Newtonian limit, speed and polarisations of gravitational waves or cosmological parameters may be determined for a generic scalar-torsion action, in analogy to a similar treatment of scalar-tensor gravity. By comparison with observations this would yield constraints on the free functions in the action. Further, foundational aspects of this class of theories may be studied, such as the number of degrees of freedom of the presence of energy conditions. We finally remark that an analogous discussion of conformal transformations, invariants and frames should also be possible for a similar class of theories in which the scalar field is non-minimally coupled to nonmetricity instead of torsion~\cite{Jarv:2018bgs}.

\begin{acknowledgments}
The author thanks Martin Kr\v{s}\v{s}\'ak and Christian Pfeifer for helpful comments and discussions. He gratefully acknowledges the full financial support of the Estonian Ministry for Education and Science through the Institutional Research Support Project IUT02-27 and Startup Research Grant PUT790, as well as the European Regional Development Fund through the Center of Excellence TK133 ``The Dark Side of the Universe''.
\end{acknowledgments}

\bibliography{scaltors}

\begin{thebibliography}{44}%
\makeatletter
\providecommand \@ifxundefined [1]{%
 \@ifx{#1\undefined}
}%
\providecommand \@ifnum [1]{%
 \ifnum #1\expandafter \@firstoftwo
 \else \expandafter \@secondoftwo
 \fi
}%
\providecommand \@ifx [1]{%
 \ifx #1\expandafter \@firstoftwo
 \else \expandafter \@secondoftwo
 \fi
}%
\providecommand \natexlab [1]{#1}%
\providecommand \enquote  [1]{``#1''}%
\providecommand \bibnamefont  [1]{#1}%
\providecommand \bibfnamefont [1]{#1}%
\providecommand \citenamefont [1]{#1}%
\providecommand \href@noop [0]{\@secondoftwo}%
\providecommand \href [0]{\begingroup \@sanitize@url \@href}%
\providecommand \@href[1]{\@@startlink{#1}\@@href}%
\providecommand \@@href[1]{\endgroup#1\@@endlink}%
\providecommand \@sanitize@url [0]{\catcode `\\12\catcode `\$12\catcode
  `\&12\catcode `\#12\catcode `\^12\catcode `\_12\catcode `\%12\relax}%
\providecommand \@@startlink[1]{}%
\providecommand \@@endlink[0]{}%
\providecommand \url  [0]{\begingroup\@sanitize@url \@url }%
\providecommand \@url [1]{\endgroup\@href {#1}{\urlprefix }}%
\providecommand \urlprefix  [0]{URL }%
\providecommand \Eprint [0]{\href }%
\providecommand \doibase [0]{http://dx.doi.org/}%
\providecommand \selectlanguage [0]{\@gobble}%
\providecommand \bibinfo  [0]{\@secondoftwo}%
\providecommand \bibfield  [0]{\@secondoftwo}%
\providecommand \translation [1]{[#1]}%
\providecommand \BibitemOpen [0]{}%
\providecommand \bibitemStop [0]{}%
\providecommand \bibitemNoStop [0]{.\EOS\space}%
\providecommand \EOS [0]{\spacefactor3000\relax}%
\providecommand \BibitemShut  [1]{\csname bibitem#1\endcsname}%
\let\auto@bib@innerbib\@empty
%</preamble>
\bibitem [{\citenamefont {Faraoni}(2004)}]{Faraoni:2004pi}%
  \BibitemOpen
  \bibfield  {author} {\bibinfo {author} {\bibfnamefont {Valerio}\ \bibnamefont
  {Faraoni}},\ }\href@noop {} {\emph {\bibinfo {title} {{Cosmology in scalar
  tensor gravity}}}}\ (\bibinfo {year} {2004})\BibitemShut {NoStop}%
%%CITATION = INSPIRE-647812;%%
\bibitem [{\citenamefont {Fujii}\ and\ \citenamefont
  {Maeda}(2007)}]{Fujii:2003pa}%
  \BibitemOpen
  \bibfield  {author} {\bibinfo {author} {\bibfnamefont {Y.}~\bibnamefont
  {Fujii}}\ and\ \bibinfo {author} {\bibfnamefont {K.}~\bibnamefont {Maeda}},\
  }\href {http://www.cambridge.org/uk/catalogue/catalogue.asp?isbn=0521811597}
  {\emph {\bibinfo {title} {{The scalar-tensor theory of gravitation}}}}\
  (\bibinfo  {publisher} {Cambridge University Press},\ \bibinfo {year}
  {2007})\BibitemShut {NoStop}%
%%CITATION = INSPIRE-618647;%%
\bibitem [{\citenamefont {Flanagan}(2004)}]{Flanagan:2004bz}%
  \BibitemOpen
  \bibfield  {author} {\bibinfo {author} {\bibfnamefont {Eanna~E.}\
  \bibnamefont {Flanagan}},\ }\bibfield  {title} {\enquote {\bibinfo {title}
  {{The Conformal frame freedom in theories of gravitation}},}\ }\href
  {\doibase 10.1088/0264-9381/21/15/N02} {\bibfield  {journal} {\bibinfo
  {journal} {Class. Quant. Grav.}\ }\textbf {\bibinfo {volume} {21}},\ \bibinfo
  {pages} {3817} (\bibinfo {year} {2004})},\ \Eprint
  {http://arxiv.org/abs/gr-qc/0403063} {arXiv:gr-qc/0403063 [gr-qc]}
  \BibitemShut {NoStop}%
%%CITATION = GR-QC/0403063;%%
\bibitem [{\citenamefont {Catena}\ \emph {et~al.}(2007)\citenamefont {Catena},
  \citenamefont {Pietroni},\ and\ \citenamefont {Scarabello}}]{Catena:2006bd}%
  \BibitemOpen
  \bibfield  {author} {\bibinfo {author} {\bibfnamefont {Riccardo}\
  \bibnamefont {Catena}}, \bibinfo {author} {\bibfnamefont {Massimo}\
  \bibnamefont {Pietroni}}, \ and\ \bibinfo {author} {\bibfnamefont {Luca}\
  \bibnamefont {Scarabello}},\ }\bibfield  {title} {\enquote {\bibinfo {title}
  {{Einstein and Jordan reconciled: a frame-invariant approach to scalar-tensor
  cosmology}},}\ }\href {\doibase 10.1103/PhysRevD.76.084039} {\bibfield
  {journal} {\bibinfo  {journal} {Phys. Rev.}\ }\textbf {\bibinfo {volume}
  {D76}},\ \bibinfo {pages} {084039} (\bibinfo {year} {2007})},\ \Eprint
  {http://arxiv.org/abs/astro-ph/0604492} {arXiv:astro-ph/0604492 [astro-ph]}
  \BibitemShut {NoStop}%
%%CITATION = ASTRO-PH/0604492;%%
\bibitem [{\citenamefont {Faraoni}\ and\ \citenamefont
  {Nadeau}(2007)}]{Faraoni:2006fx}%
  \BibitemOpen
  \bibfield  {author} {\bibinfo {author} {\bibfnamefont {Valerio}\ \bibnamefont
  {Faraoni}}\ and\ \bibinfo {author} {\bibfnamefont {Shahn}\ \bibnamefont
  {Nadeau}},\ }\bibfield  {title} {\enquote {\bibinfo {title} {{The
  (pseudo)issue of the conformal frame revisited}},}\ }\href {\doibase
  10.1103/PhysRevD.75.023501} {\bibfield  {journal} {\bibinfo  {journal} {Phys.
  Rev.}\ }\textbf {\bibinfo {volume} {D75}},\ \bibinfo {pages} {023501}
  (\bibinfo {year} {2007})},\ \Eprint {http://arxiv.org/abs/gr-qc/0612075}
  {arXiv:gr-qc/0612075 [gr-qc]} \BibitemShut {NoStop}%
%%CITATION = GR-QC/0612075;%%
\bibitem [{\citenamefont {Deruelle}\ and\ \citenamefont
  {Sasaki}(2011)}]{Deruelle:2010ht}%
  \BibitemOpen
  \bibfield  {author} {\bibinfo {author} {\bibfnamefont {Nathalie}\
  \bibnamefont {Deruelle}}\ and\ \bibinfo {author} {\bibfnamefont {Misao}\
  \bibnamefont {Sasaki}},\ }\bibfield  {title} {\enquote {\bibinfo {title}
  {{Conformal equivalence in classical gravity: the example of 'Veiled' General
  Relativity}},}\ }\bibfield  {booktitle} {\emph {\bibinfo {booktitle}
  {{Proceedings, Cosmology, the Quantum Vacuum, and Zeta Functions: Bellaterra,
  Barcelona, Spain, March 8-10, 2010}}},\ }\href {\doibase
  10.1007/978-3-642-19760-4_23} {\bibfield  {journal} {\bibinfo  {journal}
  {Springer Proc. Phys.}\ }\textbf {\bibinfo {volume} {137}},\ \bibinfo {pages}
  {247--260} (\bibinfo {year} {2011})},\ \Eprint
  {http://arxiv.org/abs/1007.3563} {arXiv:1007.3563 [gr-qc]} \BibitemShut
  {NoStop}%
%%CITATION = ARXIV:1007.3563;%%
\bibitem [{\citenamefont {Chiba}\ and\ \citenamefont
  {Yamaguchi}(2013)}]{Chiba:2013mha}%
  \BibitemOpen
  \bibfield  {author} {\bibinfo {author} {\bibfnamefont {Takeshi}\ \bibnamefont
  {Chiba}}\ and\ \bibinfo {author} {\bibfnamefont {Masahide}\ \bibnamefont
  {Yamaguchi}},\ }\bibfield  {title} {\enquote {\bibinfo {title}
  {{Conformal-Frame (In)dependence of Cosmological Observations in
  Scalar-Tensor Theory}},}\ }\href {\doibase 10.1088/1475-7516/2013/10/040}
  {\bibfield  {journal} {\bibinfo  {journal} {JCAP}\ }\textbf {\bibinfo
  {volume} {1310}},\ \bibinfo {pages} {040} (\bibinfo {year} {2013})},\ \Eprint
  {http://arxiv.org/abs/1308.1142} {arXiv:1308.1142 [gr-qc]} \BibitemShut
  {NoStop}%
%%CITATION = ARXIV:1308.1142;%%
\bibitem [{\citenamefont {Postma}\ and\ \citenamefont
  {Volponi}(2014)}]{Postma:2014vaa}%
  \BibitemOpen
  \bibfield  {author} {\bibinfo {author} {\bibfnamefont {Marieke}\ \bibnamefont
  {Postma}}\ and\ \bibinfo {author} {\bibfnamefont {Marco}\ \bibnamefont
  {Volponi}},\ }\bibfield  {title} {\enquote {\bibinfo {title} {{Equivalence of
  the Einstein and Jordan frames}},}\ }\href {\doibase
  10.1103/PhysRevD.90.103516} {\bibfield  {journal} {\bibinfo  {journal} {Phys.
  Rev.}\ }\textbf {\bibinfo {volume} {D90}},\ \bibinfo {pages} {103516}
  (\bibinfo {year} {2014})},\ \Eprint {http://arxiv.org/abs/1407.6874}
  {arXiv:1407.6874 [astro-ph.CO]} \BibitemShut {NoStop}%
%%CITATION = ARXIV:1407.6874;%%
\bibitem [{\citenamefont {Faraoni}\ \emph {et~al.}(1999)\citenamefont
  {Faraoni}, \citenamefont {Gunzig},\ and\ \citenamefont
  {Nardone}}]{Faraoni:1998qx}%
  \BibitemOpen
  \bibfield  {author} {\bibinfo {author} {\bibfnamefont {Valerio}\ \bibnamefont
  {Faraoni}}, \bibinfo {author} {\bibfnamefont {Edgard}\ \bibnamefont
  {Gunzig}}, \ and\ \bibinfo {author} {\bibfnamefont {Pasquale}\ \bibnamefont
  {Nardone}},\ }\bibfield  {title} {\enquote {\bibinfo {title} {{Conformal
  transformations in classical gravitational theories and in cosmology}},}\
  }\href@noop {} {\bibfield  {journal} {\bibinfo  {journal} {Fund. Cosmic
  Phys.}\ }\textbf {\bibinfo {volume} {20}},\ \bibinfo {pages} {121} (\bibinfo
  {year} {1999})},\ \Eprint {http://arxiv.org/abs/gr-qc/9811047}
  {arXiv:gr-qc/9811047 [gr-qc]} \BibitemShut {NoStop}%
%%CITATION = GR-QC/9811047;%%
\bibitem [{\citenamefont {Capozziello}\ \emph {et~al.}(2010)\citenamefont
  {Capozziello}, \citenamefont {Martin-Moruno},\ and\ \citenamefont
  {Rubano}}]{Capozziello:2010sc}%
  \BibitemOpen
  \bibfield  {author} {\bibinfo {author} {\bibfnamefont {S.}~\bibnamefont
  {Capozziello}}, \bibinfo {author} {\bibfnamefont {P.}~\bibnamefont
  {Martin-Moruno}}, \ and\ \bibinfo {author} {\bibfnamefont {C.}~\bibnamefont
  {Rubano}},\ }\bibfield  {title} {\enquote {\bibinfo {title} {{Physical
  non-equivalence of the Jordan and Einstein frames}},}\ }\href {\doibase
  10.1016/j.physletb.2010.04.058} {\bibfield  {journal} {\bibinfo  {journal}
  {Phys. Lett.}\ }\textbf {\bibinfo {volume} {B689}},\ \bibinfo {pages}
  {117--121} (\bibinfo {year} {2010})},\ \Eprint
  {http://arxiv.org/abs/1003.5394} {arXiv:1003.5394 [gr-qc]} \BibitemShut
  {NoStop}%
%%CITATION = ARXIV:1003.5394;%%
\bibitem [{\citenamefont {Rondeau}\ and\ \citenamefont
  {Li}(2017)}]{Rondeau:2017xck}%
  \BibitemOpen
  \bibfield  {author} {\bibinfo {author} {\bibfnamefont {François}\
  \bibnamefont {Rondeau}}\ and\ \bibinfo {author} {\bibfnamefont {Baojiu}\
  \bibnamefont {Li}},\ }\bibfield  {title} {\enquote {\bibinfo {title}
  {{Equivalence of cosmological observables in conformally related scalar
  tensor theories}},}\ }\href {\doibase 10.1103/PhysRevD.96.124009} {\bibfield
  {journal} {\bibinfo  {journal} {Phys. Rev.}\ }\textbf {\bibinfo {volume}
  {D96}},\ \bibinfo {pages} {124009} (\bibinfo {year} {2017})},\ \Eprint
  {http://arxiv.org/abs/1709.07087} {arXiv:1709.07087 [gr-qc]} \BibitemShut
  {NoStop}%
%%CITATION = ARXIV:1709.07087;%%
\bibitem [{\citenamefont {Järv}\ \emph
  {et~al.}(2015{\natexlab{a}})\citenamefont {Järv}, \citenamefont {Kuusk},
  \citenamefont {Saal},\ and\ \citenamefont {Vilson}}]{Jarv:2014hma}%
  \BibitemOpen
  \bibfield  {author} {\bibinfo {author} {\bibfnamefont {Laur}\ \bibnamefont
  {Järv}}, \bibinfo {author} {\bibfnamefont {Piret}\ \bibnamefont {Kuusk}},
  \bibinfo {author} {\bibfnamefont {Margus}\ \bibnamefont {Saal}}, \ and\
  \bibinfo {author} {\bibfnamefont {Ott}\ \bibnamefont {Vilson}},\ }\bibfield
  {title} {\enquote {\bibinfo {title} {{Invariant quantities in the
  scalar-tensor theories of gravitation}},}\ }\href {\doibase
  10.1103/PhysRevD.91.024041} {\bibfield  {journal} {\bibinfo  {journal} {Phys.
  Rev.}\ }\textbf {\bibinfo {volume} {D91}},\ \bibinfo {pages} {024041}
  (\bibinfo {year} {2015}{\natexlab{a}})},\ \Eprint
  {http://arxiv.org/abs/1411.1947} {arXiv:1411.1947 [gr-qc]} \BibitemShut
  {NoStop}%
%%CITATION = ARXIV:1411.1947;%%
\bibitem [{\citenamefont {Kuusk}\ \emph
  {et~al.}(2016{\natexlab{a}})\citenamefont {Kuusk}, \citenamefont {Jarv},\
  and\ \citenamefont {Vilson}}]{Kuusk:2015dda}%
  \BibitemOpen
  \bibfield  {author} {\bibinfo {author} {\bibfnamefont {Piret}\ \bibnamefont
  {Kuusk}}, \bibinfo {author} {\bibfnamefont {Laur}\ \bibnamefont {Jarv}}, \
  and\ \bibinfo {author} {\bibfnamefont {Ott}\ \bibnamefont {Vilson}},\
  }\bibfield  {title} {\enquote {\bibinfo {title} {{Invariant quantities in the
  multiscalar-tensor theories of gravitation}},}\ }\href {\doibase
  10.1142/S0217751X16410037} {\bibfield  {journal} {\bibinfo  {journal} {Int.
  J. Mod. Phys.}\ }\textbf {\bibinfo {volume} {A31}},\ \bibinfo {pages}
  {1641003} (\bibinfo {year} {2016}{\natexlab{a}})},\ \Eprint
  {http://arxiv.org/abs/1509.02903} {arXiv:1509.02903 [gr-qc]} \BibitemShut
  {NoStop}%
%%CITATION = ARXIV:1509.02903;%%
\bibitem [{\citenamefont {M{\o}ller}(1961)}]{Moller:1961}%
  \BibitemOpen
  \bibfield  {author} {\bibinfo {author} {\bibfnamefont {Christian}\
  \bibnamefont {M{\o}ller}},\ }\bibfield  {title} {\enquote {\bibinfo {title}
  {{Conservation Laws and Absolute Parallelism in General Relativity}},}\
  }\href@noop {} {\bibfield  {journal} {\bibinfo  {journal} {K. Dan. Vidensk.
  Selsk. Mat. Fys. Skr.}\ }\textbf {\bibinfo {volume} {1}},\ \bibinfo {pages}
  {1--50} (\bibinfo {year} {1961})}\BibitemShut {NoStop}%
\bibitem [{\citenamefont {Aldrovandi}\ and\ \citenamefont
  {Pereira}(2013)}]{Aldrovandi:2013wha}%
  \BibitemOpen
  \bibfield  {author} {\bibinfo {author} {\bibfnamefont {Ruben}\ \bibnamefont
  {Aldrovandi}}\ and\ \bibinfo {author} {\bibfnamefont {José~Geraldo}\
  \bibnamefont {Pereira}},\ }\href {\doibase 10.1007/978-94-007-5143-9} {\emph
  {\bibinfo {title} {{Teleparallel Gravity}}}},\ Vol.\ \bibinfo {volume} {173}\
  (\bibinfo  {publisher} {Springer},\ \bibinfo {address} {Dordrecht},\ \bibinfo
  {year} {2013})\BibitemShut {NoStop}%
%%CITATION = FTPHD,173,;%%
\bibitem [{\citenamefont {Maluf}(2013)}]{Maluf:2013gaa}%
  \BibitemOpen
  \bibfield  {author} {\bibinfo {author} {\bibfnamefont {J.~W.}\ \bibnamefont
  {Maluf}},\ }\bibfield  {title} {\enquote {\bibinfo {title} {{The teleparallel
  equivalent of general relativity}},}\ }\href {\doibase
  10.1002/andp.201200272} {\bibfield  {journal} {\bibinfo  {journal} {Annalen
  Phys.}\ }\textbf {\bibinfo {volume} {525}},\ \bibinfo {pages} {339--357}
  (\bibinfo {year} {2013})},\ \Eprint {http://arxiv.org/abs/1303.3897}
  {arXiv:1303.3897 [gr-qc]} \BibitemShut {NoStop}%
%%CITATION = ARXIV:1303.3897;%%
\bibitem [{\citenamefont {Golovnev}(2018)}]{Golovnev:2018red}%
  \BibitemOpen
  \bibfield  {author} {\bibinfo {author} {\bibfnamefont {Alexey}\ \bibnamefont
  {Golovnev}},\ }\bibfield  {title} {\enquote {\bibinfo {title} {{Introduction
  to teleparallel gravities}},}\ }in\ \href
  {http://inspirehep.net/record/1649207/files/arXiv:1801.06929.pdf} {\emph
  {\bibinfo {booktitle} {{9th Mathematical Physics Meeting: Summer School and
  Conference on Modern Mathematical Physics Belgrade, Serbia, September 18-23,
  2017}}}}\ (\bibinfo {year} {2018})\ \Eprint {http://arxiv.org/abs/1801.06929}
  {arXiv:1801.06929 [gr-qc]} \BibitemShut {NoStop}%
%%CITATION = ARXIV:1801.06929;%%
\bibitem [{\citenamefont {Geng}\ \emph {et~al.}(2011)\citenamefont {Geng},
  \citenamefont {Lee}, \citenamefont {Saridakis},\ and\ \citenamefont
  {Wu}}]{Geng:2011aj}%
  \BibitemOpen
  \bibfield  {author} {\bibinfo {author} {\bibfnamefont {Chao-Qiang}\
  \bibnamefont {Geng}}, \bibinfo {author} {\bibfnamefont {Chung-Chi}\
  \bibnamefont {Lee}}, \bibinfo {author} {\bibfnamefont {Emmanuel~N.}\
  \bibnamefont {Saridakis}}, \ and\ \bibinfo {author} {\bibfnamefont {Yi-Peng}\
  \bibnamefont {Wu}},\ }\bibfield  {title} {\enquote {\bibinfo {title}
  {{“Teleparallel” dark energy}},}\ }\href {\doibase
  10.1016/j.physletb.2011.09.082} {\bibfield  {journal} {\bibinfo  {journal}
  {Phys. Lett.}\ }\textbf {\bibinfo {volume} {B704}},\ \bibinfo {pages}
  {384--387} (\bibinfo {year} {2011})},\ \Eprint
  {http://arxiv.org/abs/1109.1092} {arXiv:1109.1092 [hep-th]} \BibitemShut
  {NoStop}%
%%CITATION = ARXIV:1109.1092;%%
\bibitem [{\citenamefont {Izumi}\ \emph {et~al.}(2014)\citenamefont {Izumi},
  \citenamefont {Gu},\ and\ \citenamefont {Ong}}]{Izumi:2013dca}%
  \BibitemOpen
  \bibfield  {author} {\bibinfo {author} {\bibfnamefont {Keisuke}\ \bibnamefont
  {Izumi}}, \bibinfo {author} {\bibfnamefont {Je-An}\ \bibnamefont {Gu}}, \
  and\ \bibinfo {author} {\bibfnamefont {Yen~Chin}\ \bibnamefont {Ong}},\
  }\bibfield  {title} {\enquote {\bibinfo {title} {{Acausality and Nonunique
  Evolution in Generalized Teleparallel Gravity}},}\ }\href {\doibase
  10.1103/PhysRevD.89.084025} {\bibfield  {journal} {\bibinfo  {journal} {Phys.
  Rev.}\ }\textbf {\bibinfo {volume} {D89}},\ \bibinfo {pages} {084025}
  (\bibinfo {year} {2014})},\ \Eprint {http://arxiv.org/abs/1309.6461}
  {arXiv:1309.6461 [gr-qc]} \BibitemShut {NoStop}%
%%CITATION = ARXIV:1309.6461;%%
\bibitem [{\citenamefont {Chakrabarti}\ \emph {et~al.}(2017)\citenamefont
  {Chakrabarti}, \citenamefont {Said},\ and\ \citenamefont
  {Farrugia}}]{Chakrabarti:2017moe}%
  \BibitemOpen
  \bibfield  {author} {\bibinfo {author} {\bibfnamefont {Soumya}\ \bibnamefont
  {Chakrabarti}}, \bibinfo {author} {\bibfnamefont {Jackson~Levi}\ \bibnamefont
  {Said}}, \ and\ \bibinfo {author} {\bibfnamefont {Gabriel}\ \bibnamefont
  {Farrugia}},\ }\bibfield  {title} {\enquote {\bibinfo {title} {{Some aspects
  of reconstruction using a scalar field in $f(T)$ gravity}},}\ }\href
  {\doibase 10.1140/epjc/s10052-017-5404-6} {\bibfield  {journal} {\bibinfo
  {journal} {Eur. Phys. J.}\ }\textbf {\bibinfo {volume} {C77}},\ \bibinfo
  {pages} {815} (\bibinfo {year} {2017})},\ \Eprint
  {http://arxiv.org/abs/1711.04423} {arXiv:1711.04423 [gr-qc]} \BibitemShut
  {NoStop}%
%%CITATION = ARXIV:1711.04423;%%
\bibitem [{\citenamefont {Otalora}(2013)}]{Otalora:2013tba}%
  \BibitemOpen
  \bibfield  {author} {\bibinfo {author} {\bibfnamefont {G.}~\bibnamefont
  {Otalora}},\ }\bibfield  {title} {\enquote {\bibinfo {title} {{Scaling
  attractors in interacting teleparallel dark energy}},}\ }\href {\doibase
  10.1088/1475-7516/2013/07/044} {\bibfield  {journal} {\bibinfo  {journal}
  {JCAP}\ }\textbf {\bibinfo {volume} {1307}},\ \bibinfo {pages} {044}
  (\bibinfo {year} {2013})},\ \Eprint {http://arxiv.org/abs/1305.0474}
  {arXiv:1305.0474 [gr-qc]} \BibitemShut {NoStop}%
%%CITATION = ARXIV:1305.0474;%%
\bibitem [{\citenamefont {Jamil}\ \emph {et~al.}(2012)\citenamefont {Jamil},
  \citenamefont {Momeni},\ and\ \citenamefont {Myrzakulov}}]{Jamil:2012vb}%
  \BibitemOpen
  \bibfield  {author} {\bibinfo {author} {\bibfnamefont {Mubasher}\
  \bibnamefont {Jamil}}, \bibinfo {author} {\bibfnamefont {D.}~\bibnamefont
  {Momeni}}, \ and\ \bibinfo {author} {\bibfnamefont {Ratbay}\ \bibnamefont
  {Myrzakulov}},\ }\bibfield  {title} {\enquote {\bibinfo {title} {{Stability
  of a non-minimally conformally coupled scalar field in F(T) cosmology}},}\
  }\href {\doibase 10.1140/epjc/s10052-012-2075-1} {\bibfield  {journal}
  {\bibinfo  {journal} {Eur. Phys. J.}\ }\textbf {\bibinfo {volume} {C72}},\
  \bibinfo {pages} {2075} (\bibinfo {year} {2012})},\ \Eprint
  {http://arxiv.org/abs/1208.0025} {arXiv:1208.0025 [gr-qc]} \BibitemShut
  {NoStop}%
%%CITATION = ARXIV:1208.0025;%%
\bibitem [{\citenamefont {Chen}\ \emph
  {et~al.}(2015{\natexlab{a}})\citenamefont {Chen}, \citenamefont {Wu},\ and\
  \citenamefont {Wei}}]{Chen:2014qsa}%
  \BibitemOpen
  \bibfield  {author} {\bibinfo {author} {\bibfnamefont {Zu-Cheng}\
  \bibnamefont {Chen}}, \bibinfo {author} {\bibfnamefont {You}\ \bibnamefont
  {Wu}}, \ and\ \bibinfo {author} {\bibfnamefont {Hao}\ \bibnamefont {Wei}},\
  }\bibfield  {title} {\enquote {\bibinfo {title} {{Post-Newtonian
  Approximation of Teleparallel Gravity Coupled with a Scalar Field}},}\ }\href
  {\doibase 10.1016/j.nuclphysb.2015.03.012} {\bibfield  {journal} {\bibinfo
  {journal} {Nucl. Phys.}\ }\textbf {\bibinfo {volume} {B894}},\ \bibinfo
  {pages} {422--438} (\bibinfo {year} {2015}{\natexlab{a}})},\ \Eprint
  {http://arxiv.org/abs/1410.7715} {arXiv:1410.7715 [gr-qc]} \BibitemShut
  {NoStop}%
%%CITATION = ARXIV:1410.7715;%%
\bibitem [{\citenamefont {Li}\ \emph {et~al.}(2011{\natexlab{a}})\citenamefont
  {Li}, \citenamefont {Sotiriou},\ and\ \citenamefont {Barrow}}]{Li:2010cg}%
  \BibitemOpen
  \bibfield  {author} {\bibinfo {author} {\bibfnamefont {Baojiu}\ \bibnamefont
  {Li}}, \bibinfo {author} {\bibfnamefont {Thomas~P.}\ \bibnamefont
  {Sotiriou}}, \ and\ \bibinfo {author} {\bibfnamefont {John~D.}\ \bibnamefont
  {Barrow}},\ }\bibfield  {title} {\enquote {\bibinfo {title} {{$f(T)$ gravity
  and local Lorentz invariance}},}\ }\href {\doibase
  10.1103/PhysRevD.83.064035} {\bibfield  {journal} {\bibinfo  {journal} {Phys.
  Rev.}\ }\textbf {\bibinfo {volume} {D83}},\ \bibinfo {pages} {064035}
  (\bibinfo {year} {2011}{\natexlab{a}})},\ \Eprint
  {http://arxiv.org/abs/1010.1041} {arXiv:1010.1041 [gr-qc]} \BibitemShut
  {NoStop}%
%%CITATION = ARXIV:1010.1041;%%
\bibitem [{\citenamefont {Sotiriou}\ \emph {et~al.}(2011)\citenamefont
  {Sotiriou}, \citenamefont {Li},\ and\ \citenamefont
  {Barrow}}]{Sotiriou:2010mv}%
  \BibitemOpen
  \bibfield  {author} {\bibinfo {author} {\bibfnamefont {Thomas~P.}\
  \bibnamefont {Sotiriou}}, \bibinfo {author} {\bibfnamefont {Baojiu}\
  \bibnamefont {Li}}, \ and\ \bibinfo {author} {\bibfnamefont {John~D.}\
  \bibnamefont {Barrow}},\ }\bibfield  {title} {\enquote {\bibinfo {title}
  {{Generalizations of teleparallel gravity and local Lorentz symmetry}},}\
  }\href {\doibase 10.1103/PhysRevD.83.104030} {\bibfield  {journal} {\bibinfo
  {journal} {Phys. Rev.}\ }\textbf {\bibinfo {volume} {D83}},\ \bibinfo {pages}
  {104030} (\bibinfo {year} {2011})},\ \Eprint {http://arxiv.org/abs/1012.4039}
  {arXiv:1012.4039 [gr-qc]} \BibitemShut {NoStop}%
%%CITATION = ARXIV:1012.4039;%%
\bibitem [{\citenamefont {Li}\ \emph {et~al.}(2011{\natexlab{b}})\citenamefont
  {Li}, \citenamefont {Miao},\ and\ \citenamefont {Miao}}]{Li:2011rn}%
  \BibitemOpen
  \bibfield  {author} {\bibinfo {author} {\bibfnamefont {Miao}\ \bibnamefont
  {Li}}, \bibinfo {author} {\bibfnamefont {Rong-Xin}\ \bibnamefont {Miao}}, \
  and\ \bibinfo {author} {\bibfnamefont {Yan-Gang}\ \bibnamefont {Miao}},\
  }\bibfield  {title} {\enquote {\bibinfo {title} {{Degrees of freedom of
  $f(T)$ gravity}},}\ }\href {\doibase 10.1007/JHEP07(2011)108} {\bibfield
  {journal} {\bibinfo  {journal} {JHEP}\ }\textbf {\bibinfo {volume} {07}},\
  \bibinfo {pages} {108} (\bibinfo {year} {2011}{\natexlab{b}})},\ \Eprint
  {http://arxiv.org/abs/1105.5934} {arXiv:1105.5934 [hep-th]} \BibitemShut
  {NoStop}%
%%CITATION = ARXIV:1105.5934;%%
\bibitem [{\citenamefont {Ong}\ \emph {et~al.}(2013)\citenamefont {Ong},
  \citenamefont {Izumi}, \citenamefont {Nester},\ and\ \citenamefont
  {Chen}}]{Ong:2013qja}%
  \BibitemOpen
  \bibfield  {author} {\bibinfo {author} {\bibfnamefont {Yen~Chin}\
  \bibnamefont {Ong}}, \bibinfo {author} {\bibfnamefont {Keisuke}\ \bibnamefont
  {Izumi}}, \bibinfo {author} {\bibfnamefont {James~M.}\ \bibnamefont
  {Nester}}, \ and\ \bibinfo {author} {\bibfnamefont {Pisin}\ \bibnamefont
  {Chen}},\ }\bibfield  {title} {\enquote {\bibinfo {title} {{Problems with
  Propagation and Time Evolution in f(T) Gravity}},}\ }\href {\doibase
  10.1103/PhysRevD.88.024019} {\bibfield  {journal} {\bibinfo  {journal} {Phys.
  Rev.}\ }\textbf {\bibinfo {volume} {D88}},\ \bibinfo {pages} {024019}
  (\bibinfo {year} {2013})},\ \Eprint {http://arxiv.org/abs/1303.0993}
  {arXiv:1303.0993 [gr-qc]} \BibitemShut {NoStop}%
%%CITATION = ARXIV:1303.0993;%%
\bibitem [{\citenamefont {Chen}\ \emph
  {et~al.}(2015{\natexlab{b}})\citenamefont {Chen}, \citenamefont {Izumi},
  \citenamefont {Nester},\ and\ \citenamefont {Ong}}]{Chen:2014qtl}%
  \BibitemOpen
  \bibfield  {author} {\bibinfo {author} {\bibfnamefont {Pisin}\ \bibnamefont
  {Chen}}, \bibinfo {author} {\bibfnamefont {Keisuke}\ \bibnamefont {Izumi}},
  \bibinfo {author} {\bibfnamefont {James~M.}\ \bibnamefont {Nester}}, \ and\
  \bibinfo {author} {\bibfnamefont {Yen~Chin}\ \bibnamefont {Ong}},\ }\bibfield
   {title} {\enquote {\bibinfo {title} {{Remnant Symmetry, Propagation and
  Evolution in $f$(T) Gravity}},}\ }\href {\doibase 10.1103/PhysRevD.91.064003}
  {\bibfield  {journal} {\bibinfo  {journal} {Phys. Rev.}\ }\textbf {\bibinfo
  {volume} {D91}},\ \bibinfo {pages} {064003} (\bibinfo {year}
  {2015}{\natexlab{b}})},\ \Eprint {http://arxiv.org/abs/1412.8383}
  {arXiv:1412.8383 [gr-qc]} \BibitemShut {NoStop}%
%%CITATION = ARXIV:1412.8383;%%
\bibitem [{\citenamefont {Krššák}\ and\ \citenamefont
  {Saridakis}(2016)}]{Krssak:2015oua}%
  \BibitemOpen
  \bibfield  {author} {\bibinfo {author} {\bibfnamefont {Martin}\ \bibnamefont
  {Krššák}}\ and\ \bibinfo {author} {\bibfnamefont {Emmanuel~N.}\
  \bibnamefont {Saridakis}},\ }\bibfield  {title} {\enquote {\bibinfo {title}
  {{The covariant formulation of f(T) gravity}},}\ }\href {\doibase
  10.1088/0264-9381/33/11/115009} {\bibfield  {journal} {\bibinfo  {journal}
  {Class. Quant. Grav.}\ }\textbf {\bibinfo {volume} {33}},\ \bibinfo {pages}
  {115009} (\bibinfo {year} {2016})},\ \Eprint
  {http://arxiv.org/abs/1510.08432} {arXiv:1510.08432 [gr-qc]} \BibitemShut
  {NoStop}%
%%CITATION = ARXIV:1510.08432;%%
\bibitem [{\citenamefont {Hohmann}\ \emph {et~al.}(2018)\citenamefont
  {Hohmann}, \citenamefont {Järv},\ and\ \citenamefont
  {Ualikhanova}}]{Hohmann:2018rwf}%
  \BibitemOpen
  \bibfield  {author} {\bibinfo {author} {\bibfnamefont {Manuel}\ \bibnamefont
  {Hohmann}}, \bibinfo {author} {\bibfnamefont {Laur}\ \bibnamefont {Järv}}, \
  and\ \bibinfo {author} {\bibfnamefont {Ulbossyn}\ \bibnamefont
  {Ualikhanova}},\ }\bibfield  {title} {\enquote {\bibinfo {title} {{Covariant
  formulation of scalar-torsion gravity}},}\ }\href@noop {} {\  (\bibinfo
  {year} {2018})},\ \Eprint {http://arxiv.org/abs/1801.05786} {arXiv:1801.05786
  [gr-qc]} \BibitemShut {NoStop}%
%%CITATION = ARXIV:1801.05786;%%
\bibitem [{\citenamefont {Hohmann}(2018)}]{Hohmann:2018vle}%
  \BibitemOpen
  \bibfield  {author} {\bibinfo {author} {\bibfnamefont {Manuel}\ \bibnamefont
  {Hohmann}},\ }\bibfield  {title} {\enquote {\bibinfo {title} {{Scalar-torsion
  theories of gravity I: general formalism and conformal transformations}},}\
  }\href@noop {} {\  (\bibinfo {year} {2018})},\ \Eprint
  {http://arxiv.org/abs/1801.06528} {arXiv:1801.06528 [gr-qc]} \BibitemShut
  {NoStop}%
%%CITATION = ARXIV:1801.06528;%%
\bibitem [{\citenamefont {Hohmann}\ and\ \citenamefont
  {Pfeifer}(2018)}]{Hohmann:2018dqh}%
  \BibitemOpen
  \bibfield  {author} {\bibinfo {author} {\bibfnamefont {Manuel}\ \bibnamefont
  {Hohmann}}\ and\ \bibinfo {author} {\bibfnamefont {Christian}\ \bibnamefont
  {Pfeifer}},\ }\bibfield  {title} {\enquote {\bibinfo {title} {{Scalar-torsion
  theories of gravity II: $L(T, X, Y, \phi)$ theory}},}\ }\href@noop {} {\
  (\bibinfo {year} {2018})},\ \Eprint {http://arxiv.org/abs/1801.06536}
  {arXiv:1801.06536 [gr-qc]} \BibitemShut {NoStop}%
%%CITATION = ARXIV:1801.06536;%%
\bibitem [{\citenamefont {Wright}(2016)}]{Wright:2016ayu}%
  \BibitemOpen
  \bibfield  {author} {\bibinfo {author} {\bibfnamefont {Matthew}\ \bibnamefont
  {Wright}},\ }\bibfield  {title} {\enquote {\bibinfo {title} {{Conformal
  transformations in modified teleparallel theories of gravity revisited}},}\
  }\href {\doibase 10.1103/PhysRevD.93.103002} {\bibfield  {journal} {\bibinfo
  {journal} {Phys. Rev.}\ }\textbf {\bibinfo {volume} {D93}},\ \bibinfo {pages}
  {103002} (\bibinfo {year} {2016})},\ \Eprint
  {http://arxiv.org/abs/1602.05764} {arXiv:1602.05764 [gr-qc]} \BibitemShut
  {NoStop}%
%%CITATION = ARXIV:1602.05764;%%
\bibitem [{\citenamefont {Bettoni}\ and\ \citenamefont
  {Zumalacárregui}(2015)}]{Bettoni:2015wta}%
  \BibitemOpen
  \bibfield  {author} {\bibinfo {author} {\bibfnamefont {Dario}\ \bibnamefont
  {Bettoni}}\ and\ \bibinfo {author} {\bibfnamefont {Miguel}\ \bibnamefont
  {Zumalacárregui}},\ }\bibfield  {title} {\enquote {\bibinfo {title}
  {{Kinetic mixing in scalar-tensor theories of gravity}},}\ }\href {\doibase
  10.1103/PhysRevD.91.104009} {\bibfield  {journal} {\bibinfo  {journal} {Phys.
  Rev.}\ }\textbf {\bibinfo {volume} {D91}},\ \bibinfo {pages} {104009}
  (\bibinfo {year} {2015})},\ \Eprint {http://arxiv.org/abs/1502.02666}
  {arXiv:1502.02666 [gr-qc]} \BibitemShut {NoStop}%
%%CITATION = ARXIV:1502.02666;%%
\bibitem [{\citenamefont {Järv}\ \emph
  {et~al.}(2015{\natexlab{b}})\citenamefont {Järv}, \citenamefont {Kuusk},
  \citenamefont {Saal},\ and\ \citenamefont {Vilson}}]{Jarv:2015kga}%
  \BibitemOpen
  \bibfield  {author} {\bibinfo {author} {\bibfnamefont {Laur}\ \bibnamefont
  {Järv}}, \bibinfo {author} {\bibfnamefont {Piret}\ \bibnamefont {Kuusk}},
  \bibinfo {author} {\bibfnamefont {Margus}\ \bibnamefont {Saal}}, \ and\
  \bibinfo {author} {\bibfnamefont {Ott}\ \bibnamefont {Vilson}},\ }\bibfield
  {title} {\enquote {\bibinfo {title} {{Transformation properties and general
  relativity regime in scalar–tensor theories}},}\ }\href {\doibase
  10.1088/0264-9381/32/23/235013} {\bibfield  {journal} {\bibinfo  {journal}
  {Class. Quant. Grav.}\ }\textbf {\bibinfo {volume} {32}},\ \bibinfo {pages}
  {235013} (\bibinfo {year} {2015}{\natexlab{b}})},\ \Eprint
  {http://arxiv.org/abs/1504.02686} {arXiv:1504.02686 [gr-qc]} \BibitemShut
  {NoStop}%
%%CITATION = ARXIV:1504.02686;%%
\bibitem [{\citenamefont {Kuusk}\ \emph
  {et~al.}(2016{\natexlab{b}})\citenamefont {Kuusk}, \citenamefont {Rünkla},
  \citenamefont {Saal},\ and\ \citenamefont {Vilson}}]{Kuusk:2016rso}%
  \BibitemOpen
  \bibfield  {author} {\bibinfo {author} {\bibfnamefont {Piret}\ \bibnamefont
  {Kuusk}}, \bibinfo {author} {\bibfnamefont {Mihkel}\ \bibnamefont {Rünkla}},
  \bibinfo {author} {\bibfnamefont {Margus}\ \bibnamefont {Saal}}, \ and\
  \bibinfo {author} {\bibfnamefont {Ott}\ \bibnamefont {Vilson}},\ }\bibfield
  {title} {\enquote {\bibinfo {title} {{Invariant slow-roll parameters in
  scalar–tensor theories}},}\ }\href {\doibase
  10.1088/0264-9381/33/19/195008} {\bibfield  {journal} {\bibinfo  {journal}
  {Class. Quant. Grav.}\ }\textbf {\bibinfo {volume} {33}},\ \bibinfo {pages}
  {195008} (\bibinfo {year} {2016}{\natexlab{b}})},\ \Eprint
  {http://arxiv.org/abs/1605.07033} {arXiv:1605.07033 [gr-qc]} \BibitemShut
  {NoStop}%
%%CITATION = ARXIV:1605.07033;%%
\bibitem [{\citenamefont {Järv}\ \emph {et~al.}(2017)\citenamefont {Järv},
  \citenamefont {Kannike}, \citenamefont {Marzola}, \citenamefont {Racioppi},
  \citenamefont {Raidal}, \citenamefont {Rünkla}, \citenamefont {Saal},\ and\
  \citenamefont {Veermäe}}]{Jarv:2016sow}%
  \BibitemOpen
  \bibfield  {author} {\bibinfo {author} {\bibfnamefont {Laur}\ \bibnamefont
  {Järv}}, \bibinfo {author} {\bibfnamefont {Kristjan}\ \bibnamefont
  {Kannike}}, \bibinfo {author} {\bibfnamefont {Luca}\ \bibnamefont {Marzola}},
  \bibinfo {author} {\bibfnamefont {Antonio}\ \bibnamefont {Racioppi}},
  \bibinfo {author} {\bibfnamefont {Martti}\ \bibnamefont {Raidal}}, \bibinfo
  {author} {\bibfnamefont {Mihkel}\ \bibnamefont {Rünkla}}, \bibinfo {author}
  {\bibfnamefont {Margus}\ \bibnamefont {Saal}}, \ and\ \bibinfo {author}
  {\bibfnamefont {Hardi}\ \bibnamefont {Veermäe}},\ }\bibfield  {title}
  {\enquote {\bibinfo {title} {{Frame-Independent Classification of
  Single-Field Inflationary Models}},}\ }\href {\doibase
  10.1103/PhysRevLett.118.151302} {\bibfield  {journal} {\bibinfo  {journal}
  {Phys. Rev. Lett.}\ }\textbf {\bibinfo {volume} {118}},\ \bibinfo {pages}
  {151302} (\bibinfo {year} {2017})},\ \Eprint
  {http://arxiv.org/abs/1612.06863} {arXiv:1612.06863 [hep-ph]} \BibitemShut
  {NoStop}%
%%CITATION = ARXIV:1612.06863;%%
\bibitem [{\citenamefont {Karam}\ \emph {et~al.}(2017)\citenamefont {Karam},
  \citenamefont {Pappas},\ and\ \citenamefont {Tamvakis}}]{Karam:2017zno}%
  \BibitemOpen
  \bibfield  {author} {\bibinfo {author} {\bibfnamefont {Alexandros}\
  \bibnamefont {Karam}}, \bibinfo {author} {\bibfnamefont {Thomas}\
  \bibnamefont {Pappas}}, \ and\ \bibinfo {author} {\bibfnamefont {Kyriakos}\
  \bibnamefont {Tamvakis}},\ }\bibfield  {title} {\enquote {\bibinfo {title}
  {{Frame-dependence of higher-order inflationary observables in scalar-tensor
  theories}},}\ }\href {\doibase 10.1103/PhysRevD.96.064036} {\bibfield
  {journal} {\bibinfo  {journal} {Phys. Rev.}\ }\textbf {\bibinfo {volume}
  {D96}},\ \bibinfo {pages} {064036} (\bibinfo {year} {2017})},\ \Eprint
  {http://arxiv.org/abs/1707.00984} {arXiv:1707.00984 [gr-qc]} \BibitemShut
  {NoStop}%
%%CITATION = ARXIV:1707.00984;%%
\bibitem [{\citenamefont {Bahamonde}\ and\ \citenamefont
  {Wright}(2015)}]{Bahamonde:2015hza}%
  \BibitemOpen
  \bibfield  {author} {\bibinfo {author} {\bibfnamefont {Sebastian}\
  \bibnamefont {Bahamonde}}\ and\ \bibinfo {author} {\bibfnamefont {Matthew}\
  \bibnamefont {Wright}},\ }\bibfield  {title} {\enquote {\bibinfo {title}
  {{Teleparallel quintessence with a nonminimal coupling to a boundary
  term}},}\ }\href {\doibase 10.1103/PhysRevD.92.084034,
  10.1103/PhysRevD.93.109901} {\bibfield  {journal} {\bibinfo  {journal} {Phys.
  Rev.}\ }\textbf {\bibinfo {volume} {D92}},\ \bibinfo {pages} {084034}
  (\bibinfo {year} {2015})},\ \bibinfo {note} {[Erratum: Phys.
  Rev.D93,no.10,109901(2016)]},\ \Eprint {http://arxiv.org/abs/1508.06580}
  {arXiv:1508.06580 [gr-qc]} \BibitemShut {NoStop}%
%%CITATION = ARXIV:1508.06580;%%
\bibitem [{\citenamefont {Bahamonde}\ \emph {et~al.}(2018)\citenamefont
  {Bahamonde}, \citenamefont {Marciu},\ and\ \citenamefont
  {Rudra}}]{Bahamonde:2018miw}%
  \BibitemOpen
  \bibfield  {author} {\bibinfo {author} {\bibfnamefont {Sebastian}\
  \bibnamefont {Bahamonde}}, \bibinfo {author} {\bibfnamefont {Mihai}\
  \bibnamefont {Marciu}}, \ and\ \bibinfo {author} {\bibfnamefont {Prabir}\
  \bibnamefont {Rudra}},\ }\bibfield  {title} {\enquote {\bibinfo {title}
  {{Generalised teleparallel quintom dark energy non-minimally coupled with the
  scalar torsion and a boundary term}},}\ }\href@noop {} {\  (\bibinfo {year}
  {2018})},\ \Eprint {http://arxiv.org/abs/1802.09155} {arXiv:1802.09155
  [gr-qc]} \BibitemShut {NoStop}%
%%CITATION = ARXIV:1802.09155;%%
\bibitem [{\citenamefont {Damour}\ and\ \citenamefont
  {Esposito-Farese}(1992)}]{Damour:1992we}%
  \BibitemOpen
  \bibfield  {author} {\bibinfo {author} {\bibfnamefont {Thibault}\
  \bibnamefont {Damour}}\ and\ \bibinfo {author} {\bibfnamefont {Gilles}\
  \bibnamefont {Esposito-Farese}},\ }\bibfield  {title} {\enquote {\bibinfo
  {title} {{Tensor multiscalar theories of gravitation}},}\ }\href {\doibase
  10.1088/0264-9381/9/9/015} {\bibfield  {journal} {\bibinfo  {journal} {Class.
  Quant. Grav.}\ }\textbf {\bibinfo {volume} {9}},\ \bibinfo {pages}
  {2093--2176} (\bibinfo {year} {1992})}\BibitemShut {NoStop}%
%%CITATION = CQGRD,9,2093;%%
\bibitem [{\citenamefont {Berkin}\ and\ \citenamefont
  {Hellings}(1994)}]{Berkin:1993bt}%
  \BibitemOpen
  \bibfield  {author} {\bibinfo {author} {\bibfnamefont {Andrew~L.}\
  \bibnamefont {Berkin}}\ and\ \bibinfo {author} {\bibfnamefont {Ronald~W.}\
  \bibnamefont {Hellings}},\ }\bibfield  {title} {\enquote {\bibinfo {title}
  {{Multiple field scalar - tensor theories of gravity and cosmology}},}\
  }\href {\doibase 10.1103/PhysRevD.49.6442} {\bibfield  {journal} {\bibinfo
  {journal} {Phys. Rev.}\ }\textbf {\bibinfo {volume} {D49}},\ \bibinfo {pages}
  {6442--6449} (\bibinfo {year} {1994})},\ \Eprint
  {http://arxiv.org/abs/gr-qc/9401033} {arXiv:gr-qc/9401033 [gr-qc]}
  \BibitemShut {NoStop}%
%%CITATION = GR-QC/9401033;%%
\bibitem [{\citenamefont {Hohmann}\ \emph {et~al.}(2016)\citenamefont
  {Hohmann}, \citenamefont {Jarv}, \citenamefont {Kuusk}, \citenamefont
  {Randla},\ and\ \citenamefont {Vilson}}]{Hohmann:2016yfd}%
  \BibitemOpen
  \bibfield  {author} {\bibinfo {author} {\bibfnamefont {Manuel}\ \bibnamefont
  {Hohmann}}, \bibinfo {author} {\bibfnamefont {Laur}\ \bibnamefont {Jarv}},
  \bibinfo {author} {\bibfnamefont {Piret}\ \bibnamefont {Kuusk}}, \bibinfo
  {author} {\bibfnamefont {Erik}\ \bibnamefont {Randla}}, \ and\ \bibinfo
  {author} {\bibfnamefont {Ott}\ \bibnamefont {Vilson}},\ }\bibfield  {title}
  {\enquote {\bibinfo {title} {{Post-Newtonian parameter $\gamma$ for
  multiscalar-tensor gravity with a general potential}},}\ }\href {\doibase
  10.1103/PhysRevD.94.124015} {\bibfield  {journal} {\bibinfo  {journal} {Phys.
  Rev.}\ }\textbf {\bibinfo {volume} {D94}},\ \bibinfo {pages} {124015}
  (\bibinfo {year} {2016})},\ \Eprint {http://arxiv.org/abs/1607.02356}
  {arXiv:1607.02356 [gr-qc]} \BibitemShut {NoStop}%
%%CITATION = ARXIV:1607.02356;%%
\bibitem [{\citenamefont {Järv}\ \emph {et~al.}(2018)\citenamefont {Järv},
  \citenamefont {Rünkla}, \citenamefont {Saal},\ and\ \citenamefont
  {Vilson}}]{Jarv:2018bgs}%
  \BibitemOpen
  \bibfield  {author} {\bibinfo {author} {\bibfnamefont {Laur}\ \bibnamefont
  {Järv}}, \bibinfo {author} {\bibfnamefont {Mihkel}\ \bibnamefont {Rünkla}},
  \bibinfo {author} {\bibfnamefont {Margus}\ \bibnamefont {Saal}}, \ and\
  \bibinfo {author} {\bibfnamefont {Ott}\ \bibnamefont {Vilson}},\ }\bibfield
  {title} {\enquote {\bibinfo {title} {{Nonmetricity formulation of general
  relativity and its scalar-tensor extension}},}\ }\href@noop {} {\  (\bibinfo
  {year} {2018})},\ \Eprint {http://arxiv.org/abs/1802.00492} {arXiv:1802.00492
  [gr-qc]} \BibitemShut {NoStop}%
%%CITATION = ARXIV:1802.00492;%%
\end{thebibliography}%
\end{document}